\documentclass[trackchanges,twocolumn]{aastex7}
\usepackage{graphicx}
\usepackage{amsmath}
\usepackage{booktabs}
\usepackage{CJK}

\begin{document}

\begin{CJK*}{UTF8}{gbsn}

\title{Probing Red Giant Interiors with G-Dominated Mixed Modes \uppercase\expandafter{\romannumeral1}: The Cases of KIC 9145955, KIC 9970396, KIC 9882316 and KIC 11968334
}

\correspondingauthor{Xinyi Zhang, Tanda Li, Shaolan Bi}
\email{zhangxy$\_$cn@bnu.edu.cn, litanda@bnu.edu.cn, bisl@bnu.edu.cn}

\author[0000-0003-1860-1851]{Xinyi Zhang} 
\affiliation{Institute for Frontiers in Astronomy and Astrophysics,\\
 Beijing Normal University, \\
 Beijing 102206, People{'}s Republic of China}
\affiliation{School of Physics and Astronomy,
 Beijing Normal University, Beijing 100875,
 People's Republic of China}
 \email{zhangxy\_cn@bnu.edu.cn}

\author[0000-0001-6396-2563] {Tanda Li}
\affiliation{Institute for Frontiers in Astronomy and Astrophysics,\\
 Beijing Normal University, \\
 Beijing 102206, People{'}s Republic of China} 
\affiliation{School of Physics and Astronomy,
 Beijing Normal University, Beijing 100875,
 People's Republic of China}
\email{litanda@bnu.edu.cn}

\author[0000-0002-8004-549X] {Jianxing Chen}
\affiliation{School of Physics and Astronomy,
 Beijing Normal University, Beijing 100875,
 People's Republic of China}
\email{jxchen@bnu.edu.cn}

\author[0000-0003-3112-1967] {Xinghao Chen}
\affiliation{Yunnan Observatories, Chinese Academy of Sciences,\\
PO Box 110, Kunming 650216, People's Republic of China} 
\email{chenxinghao1003@163.com} 

\author[0000-0001-9313-251X]{Gang Li}
\affiliation{Institute of Astronomy (IvS), Department of Physics and Astronomy, KU Leuven, Celestijnenlaan 200D, 3001 Leuven, Belgium}
\email{gang.li@kuleuven.be}

\author[0000-0002-6647-3957] {Chunqian Li} 
\affiliation{Institute for Frontiers in Astronomy and Astrophysics,\\
 Beijing Normal University, \\
 Beijing 102206, People{'}s Republic of China} 
\affiliation{School of Physics and Astronomy,
 Beijing Normal University, Beijing 100875,
 People's Republic of China}
\email{lcq@bnu.edu.cn}

\author[0000-0002-7642-7583] {Shaolan Bi} 
\affiliation{Institute for Frontiers in Astronomy and Astrophysics,\\
 Beijing Normal University, \\
 Beijing 102206, People{'}s Republic of China} 
\affiliation{School of Physics and Astronomy,
 Beijing Normal University, Beijing 100875,
 People's Republic of China}
\email{bisl@bnu.edu.cn}

\begin{abstract}

We perform a detailed asteroseismic analysis of four red giants observed by Kepler: KIC 9145955, KIC 9970396, KIC 9882316, and KIC 11968334. Our study is based on individual oscillation frequencies, with particular emphasis on gravity-dominated (g-dominated) mixed modes. These modes are highly sensitive to the deep stellar interior and serve as powerful diagnostics of core structure, convective overshooting, and internal rotation. Moreover, surface effects have minimal impact on g-dominated mixed modes. To ensure accurate frequency matching between observations and theoretical models, we apply a mode-identification technique that effectively distinguishes p-dominated from g-dominated modes. Although a definitive confirmation of this trend requires a substantially larger asteroseismic sample, our best-fitting models suggest that the derived convective overshooting parameter ($f_{ov}$) increases with stellar mass. In particular, within our sample the star with a mass exceeding $1.4M_{\odot}$ requires $f_{ov} > 0.01$, whereas lower-mass red giants tend to have $f_{ov}$ <0.01. In addition, the average core rotation rate of KIC 11968334 is precisely determined to be $0.7409\pm0.0113 \mu$Hz from the asteroseismic model.

\end{abstract}



\section{Introduction} \label{sec1}

Over the past decades, asteroseismology has significantly improved our understanding of stellar interiors and evolution, largely thanks to space missions such as CoRoT \citep{2006ESASP1306...33B}, Kepler/K2 \citep{2009IAUS..253..289B,2014PASP..126..398H}, and TESS \citep{2014SPIE.9143E..20R}. The long-duration, high-precision light curves from these missions have enabled extensive detections of mixed modes in red giant stars \citep{2010ApJ...713L.176B,2011Sci...332..205B,Mosser2011a}. These modes, which are most prominent in dipole ($l=1$) oscillations, arise from the coupling between the pressure (p-) mode cavity in the envelope and the gravity (g-) mode cavity in the core \citep{2010aste.book.....A}, making them powerful probes of stellar interiors. For example, the period spacing of gravity-dominated (g-dominated) mixed modes distinguishes between stars ascending the red giant branch (RGB) and those in the red clump (RC) \citep{2011Natur.471..608B,2014A&A...572L...5M}. Furthermore, mixed modes offer critical insights into internal rotation through the analysis of rotational frequency splittings. Stellar rotation lifts the degeneracy of non-radial modes, splitting a single frequency into multiple components corresponding to prograde and retrograde traveling waves. Analysis of these splittings in red giants has revealed significant radial differential rotation, with stellar cores rotating much faster than their envelopes \citep{2012Natur.481...55B,2012ApJ...756...19D,Mosser2012a}.

By modeling individual oscillation frequencies, stellar parameters can be inferred with a precision that often exceeds other observational methods (see \citealp{2021RvMP...93a5001A} for a review). This detailed asteroseismic modeling has been successfully applied to numerous red giants\citep{Perez2016,2018ApJ...855...16Z,2018MNRAS.475..981L,2020MNRAS.494..511Z,2021MNRAS.505.2336M,2022ApJ...927..167L,2023ApJ...953..182W}. Despite these successes, several challenges persist. One major issue is the surface effect, which refers to systematic frequency offsets between models and observations. This is caused by an incomplete treatment of near-surface convection \citep{1988Natur.336..634C,1996Sci...272.1286C,1997MNRAS.284..527C}. Furthermore, uncertainties in input physics, particularly the mixing-length parameter \citep{2018MNRAS.475..981L} and convective core overshooting \citep{2017MNRAS.472.4900C,2018A&A...614A.128P}, can introduce systematic errors into the inferred stellar parameters.

Mixed modes are less sensitive to the surface effect than pure p modes due to their high mode inertia \citep{2008ApJ...683L.175K,2018ApJ...855...16Z,2020MNRAS.494..511Z}. Moreover, g-dominated mixed modes are highly sensitive to core properties and can provide valuable constraints on the convective core overshooting parameter $f_{ov}$ in post-main-sequence stars \citep{2013ApJ...766..118M,2021A&A...647A.187N}. This parameter is particularly crucial because the extent of overshooting determines the size of the helium core formed during the main sequence and affects the age of stars on the red giant branch \citep{2013sse..book.....K}. Several studies employing different techniques have found evidence that $f_{ov}$ increases with stellar mass\citep{2006ApJS..162..375V,2007A&A...475.1019C,2012MNRAS.427..127B,2016A&A...589A..93D,2017ApJ...849...18C,2018ApJ...859..100C,2019ApJ...876..134C,2018ApJ...856..125H,2022MNRAS.512.4852Z}. , although the precise form of this dependence and its behavior at low masses remain subjects of ongoing investigation. In this context, detailed asteroseismic modeling of individual red giants offers a complementary and independent approach to constraining $f_{ov}$.
Furthermore, our previous work \citep{Zhang2022ApJ} demonstrated that g-dominated mixed modes are highly effective for probing helium core properties and precisely constraining the ages of red giants.

Building on these results, this study aims to derive the fundamental parameters for four specific Kepler red giants: KIC 9145955, KIC 9970396, KIC 9882316, and KIC 11968334 by using g-dominated mixed modes as observation constraints. In addition to determining stellar parameters, we investigate the mass dependence of the convective core overshooting parameter. For KIC 11968334, we further analyze rotational splittings in dipole ($l=1$) modes to estimate the average core and envelope rotation rates. This paper is organized as follows: Section 2 presents the photometric and spectroscopic data for the target stars. Section 3 details our asteroseismic modeling method. In Section 4, we describe the selection of the best-fitting models (Section 4.1), analyze the period spacing for KIC 9970396 (Section 4.2), address the surface effect (Section 4.3), investigate the overshooting parameter (Section 4.4), discuss the mixing-length parameter (Section 4.5), and present the rotation analysis (Section 4.6). Finally, we summarize our findings in Section 5.


\section{Photometric and Spectroscopic observations} \label{sec2}

We selected four \textit{Kepler} red giants for asteroseismic analysis: KIC 9145955, KIC 9970396, KIC 9882316, and KIC 11968334. Additionally, all four targets are part of the well-studied APOKASC sample, which has been extensively investigated by various groups. For instance, the evolutionary stages of these stars have been systematically evaluated in previous literature (e.g., \citealt{2019MNRAS.489.4641E}), and robust asteroseismic constraints have been provided through detailed peak-bagging analyses (e.g., \citealt{2019arXiv190609428K}). Building upon this rich contextual background, we adopt the specific individual oscillation frequencies derived in several dedicated studies for our detailed modeling. The individual oscillation frequencies of KIC 9145955 were extracted by \citet{2018ApJ...855...16Z} and are adopted in this study. For KIC 9970396, which was identified as a pulsating red giant in a detached eclipsing binary by \citet{2013ApJ...767...82G,2014ApJ...785....5G}, we use the frequencies reported by \citet{2018MNRAS.475..981L}. The frequencies of KIC 9882316 and KIC 11968334 are taken from \citet{corsaro2015}. All four stars exhibit mixed modes, and rotational splitting is detected in KIC 11968334. The asteroseismic and spectroscopic parameters used in our analysis are listed in Table \ref{t1}, and the individual oscillation frequencies are provided in Tables \ref{ta1} - \ref{ta4}. $\Delta\nu$ and $\Delta P_{obs}$ in Table \ref{t1} represent observational large frequency separation and period spacing, respectively.

\begin{deluxetable*}{lcccc}
\tablecaption{Stellar Parameters of Four Red Giants}
\label{t1}
\tablehead{
\colhead{Parameter} & \colhead{KIC 9145955} & \colhead{KIC 9970396} & \colhead{KIC 9882316} & \colhead{KIC 11968334}
}
\startdata
$T_{\rm eff}$ (K)   &$4943^{a}$                &$4916\pm68^{h}$             &$5093\pm91^{b}$         &$4826\pm91^{b}$ \\
                    &$4925\pm91^{b}$           &$4789\pm91^{i}$             &$5140.354\pm14.453^{c}$ &$4765.037\pm7.597^{c}$\\
                    &$4989.284\pm10.246^{c}$   &$4789.8154\pm8.600^{c}$     &                        &$4709.92\pm42.56^{n}$  \\
                    &$4995.85\pm86.27^{n}$     &$4868.11\pm23.40^{n}$       &                        &\\
log $g$ (dex)       &$2.85^{a}$                &$3.1\pm0.1^{h}$             &$3.20\pm0.11^{b}$       &$3.01\pm0.11^{b}$ \\
                    &$3.04\pm0.11^{b}$         &$2.7\pm0.1^{i}$             &$3.158\pm0.031^{c}$     &$3.060\pm0.020^{c}$ \\
                    &$3.066\pm0.025^{c}$       &$2.691\pm0.024^{c}$         &                        &$3.166\pm0.070^{n}$ \\
                    &$3.104\pm0.142^{n}$       &$2.852\pm0.037^{n}$         &                        &\\
$ \rm[Fe/H]$        &$-0.32\pm0.03^{b}$        &$-0.245\pm0.007^{c}$        &$-0.41\pm0.04^{b}$    &$0.35\pm0.03^{b}$\\               
                    &$-0.341\pm0.007^{c}$      &$-0.23\pm0.03^{h}$          &$-0.410\pm0.009^{c}$  &$0.174 \pm 0.007^{c}$\\
                    &$-0.294\pm0.083^{n}$      &$-0.18\pm0.07^{i}$          &                      &$0.210\pm0.044^{n}$\\
                    &                          &$-0.244\pm0.020^{n}$        &   &\\
$\Delta\nu$ ($\mu$Hz)   &$11.00^{a}$           &$6.3^{j}$               &$13.78\pm0.07^{b}$   &$11.41\pm0.06^{b}$ \\
                        &$11.00\pm0.06^{b}$    &$6.336\pm0.005^{k}$     &          & \\
                        &$11.065^{d}$          &$6.320\pm0.010^{h}$     &          & \\
                        &                      &$6.30\pm0.01^{l}$       &          & \\
$\nu_{\rm max}$ ($\mu$Hz) &$130.0^{a}$         &$63.9^{j}$              &$182.0\pm0.5^{b}$      &$141.4\pm0.3^{b}$ \\
                          &$131.7\pm0.2^{b}$   &$63.19\pm0.23^{k}$      &                       & \\
                          &                    &$63.70\pm0.16^{h}$      &                       & \\
                          &                    &$63.8\pm0.5^{l}$        &                       & \\
$\Delta \Pi_{\rm 1,obs}$ (s)    &$77.01^{a}$         &$64\pm0.06^{o}$         &$80.59\pm0.02^{g}$     &$78.10\pm0.10^{g}$ \\
                                &$76.98\pm0.03^{e}$  &$56.73^{m}$             &                       & \\
                                &$77.1^{f}$          &                        &                       & \\
                                &$77.01^{d}$         &                        &                       & \\
                                &$77.01\pm0.02^{g}$  &                        &                       &\\
\enddata
\tablenotetext{a}{\citet{2016MNRAS.457.4454T}}
\tablenotetext{b}{\citet{Perez2016}}
\tablenotetext{c}{\citet{2022ApJS..259...35A}(APOGEE)}
\tablenotetext{d}{\citet{2018ApJ...855...16Z}}
\tablenotetext{e}{\citet{2015MNRAS.447.1935D}}
\tablenotetext{f}{\citet{Vrard2016}}
\tablenotetext{g}{\citet{Mosser2012b}}
\tablenotetext{h}{\citet{2016ApJ...832..121G}(ACRES)}
\tablenotetext{i}{\citet{2015ApJS..219...12A}(APOGEE)}
\tablenotetext{j}{\citet{2013ApJ...767...82G}}
\tablenotetext{k}{\citet{2014ApJ...785....5G}}
\tablenotetext{l}{\citet{2018MNRAS.475..981L}}
\tablenotetext{m}{\citet{Zhang2022ApJ}}
\tablenotetext{n}{\citet{2018ApJS..238...30Z}(LAMOST)}
\tablenotetext{o}{this work}
\end{deluxetable*}

\section{Asteroseismic Model}\label{sec3}
\subsection{Input Parameters}\label{s3.1}

This work employs the one-dimensional stellar evolution code, Modules for Experiments in Stellar Astrophysics (MESA, version 10398; \citealp{2011ApJS..192....3P,2013ApJS..208....4P,2015ApJS..220...15P,2018ApJS..234...34P}, to generate theoretical models. MESA is an open-source tool that integrates various physical modules and is widely used in stellar astrophysics. In this study, we adopt the  ``pulse\_adipls'' module developed by Christensen-Dalsgaard (2008) to compute the evolutionary models and their corresponding adiabatic oscillation frequencies. For the equation of state (EOS), we use the OPAL EOS tables provided by \citet{2002ApJ...576.1064R}. The opacity tables are taken from \citet{1996ApJ...464..943I} for high-temperature regions and \citet{2005ApJ...623..585F} for low-temperature regions. The solar composition GS98 \citep{1998SSRv...85..161G} is used as the initial composition
in metal. We adopt the mixing-length theory (MLT) of \citet{1958ZA.....46..108B} to deal with stellar convection. The convective overshooting in the core is explained with the theory of \citet{2000A&A...360..952H}. The overshoot mixing diffusion coefficient $D_{ov}$ is defined as

\begin{equation}
D_{\rm ov}=D_{\rm conv,0} exp\left (\frac{-2z}{f_{\rm ov}H_{\rm p,0}} \right ),
\label{e1}
\end{equation}

where $f_{ov}$ is an adjustable parameter, $D_{\rm conv,0}$ is the MLT derived diffusion coefficient at a user-defined location near the Schwarzschild boundary, $H_{\rm p,0}$ is the pressure scale height at that location, and $z$ is the distance in the radiative layer away from that location, for more details, see \citealp{2018ApJ...855...16Z,2020MNRAS.494..511Z}. The treatment of the stellar atmosphere differs slightly between models. For KIC 9145955 and KIC 9970396, we use the Eddington gray atmosphere to determine surface boundary conditions. For KIC 9882316 and KIC 11968334, we use a simple photosphere atmospheric model because the Eddington model in the MESA 10398 version cannot generate pre-main-sequence models for some of the parameters in Table \ref{t2}. In Section 4.5, we discuss how varying the mixing-length parameter affects the fundamental parameters of red giants when the photosphere atmospheric model serves as the boundary condition.

\subsection{Model Grids}\label{s3.2}

Table \ref{t2} summarizes the input parameters used for the four red giant stars: KIC 9145955, KIC 9970396, KIC 9882316, and KIC 11968334. Each theoretical model evolves from the pre-main-sequence to the red giant stage, with the initial helium mass fraction $Y$ set by the relation $Y=0.245+1.54Z$ \citep{2008ApJS..178...89D,2014AJ....148...85T}. The parameter ranges of the initial masses and metallicity of the four red giants in Table 2 refer to the work of \citet{Perez2016} and the observed values of $\rm [Fe/H]$ provided in Table \ref{t1}.

The mixing-length parameter $\alpha$ is an important factor in red giant modeling. In our previous work \citep{2018ApJ...855...16Z,2020MNRAS.494..511Z}, asteroseismic modeling of KIC 9145955 and KIC 9970396 constrained $\alpha$ to be approximately 2.0. This value is also consistent with Li et al. (2018), who calibrated $\alpha$ for red giants in detached eclipsing binaries with a simple photospheric boundary condition. Moreover, our previous work \citep{2018ApJ...855...16Z,2020MNRAS.494..511Z} showed that helium-core mass estimates derived from mixed modes are relatively insensitive to $\alpha$, supporting the choice of $\alpha=2.0$ for KIC 9882316 and KIC 11968334 as well. Based on this calibration, we adopt a fixed value of $\alpha$=2.0 for all models in Grid A of Table \ref{t2}. 

In Grid B of Table \ref{t2}, KIC 9882316 serves as a representative case to investigate the impact of varying mixing-length parameters on the inferred stellar properties. Beyond the mixing-length parameter $\alpha$, the convective core overshooting parameter $f_{ov}$ is another key parameter. We first perform a broad parameter search using Grid A, and then use these results to refine the overshooting range explored in Grid B. The details of this procedure are described in Sections \ref{s4.4} and \ref{s4.5}.

\begin{table*}
\caption{\label{t2}
Initial Input Parameters of Theoretical Models.}
\centering
\begin{tabular}{ccccc}
\hline\hline
				&mass M/$\rm \delta M$	&metallicity Z/$\rm \delta Z$	&Mixing-length parameter $\alpha$\ /$\delta \alpha$ 		&core overshooting parameter $f_{ov}$ / $\delta f_{ov}$\\
				&($\rm M_{\odot}$)\\
\hline
&&&Grid A\\
\hline
KIC 9145955		&1.18-1.30/0.01			&0.006-0.015/0.001		&2.0				&0.000-0.020/0.001\\
KIC 9970396		&1.11-1.20/0.01			&0.007-0.014/0.001		&2.0				&0.000-0.020/0.001\\
KIC 9882316		&1.35-1.43/0.01			&0.001-0.009/0.001		&2.0				&0.000-0.020/0.001\\
KIC 11968334	&1.31-1.40/0.01			&0.020-0.032/0.001	    &2.0				&0.000-0.020/0.001\\
\hline
&&&Grid B\\
KIC 9882316		&1.35-1.43/0.01			&0.003-0.009/0.001		&1.7-2.4/0.1		&0.010-0.015/0.001\\
\hline
\end{tabular}
\end{table*}

\subsection{Model Fittings}\label{s3.3}

During the red giant stage, mixed modes are predominantly observed in $l=1$ modes. In general, each radial order typically contains only a few p-dominated mixed modes, but several g-dominated mixed modes \citep{Mosser2012b}. 
In our previous work, we developed a mode identification method that effectively distinguishes between p-dominated and g-dominated mixed modes. This method identifies both the p-dominated and g-dominated $l=1$ modes in both the observed and theoretical frequency spectra. Detailed mode identification for KIC 9145955 and KIC 9970396 was presented in our previous work \citep[see][]{2018ApJ...855...16Z,2020MNRAS.494..511Z}. Based on this foundation, we apply the same procedure to KIC 9882316 and KIC 11968334 and provide a brief summary here. The pure p-mode eigenfrequencies are computed using the asymptotic relation proposed by \citet{Mosser2011b,Mosser2012b}:

\begin{equation}
\nu_{n_{p},l}=\left [ n_{p}+\frac{l}{2}+\epsilon (\Delta \nu)-d_{0l}+\frac{\alpha (\Delta \nu)}{2}[n_{p}-n_{max}] ^2\right ]\Delta \nu,
\label{e2}
\end{equation}

where $n_{p}$ is the p-mode radial order, $l$ is the angular degree, $\epsilon$ is the phase offset, $d_{0l}$ accounts for the small separation, $\alpha$ is a small constant and $n_{max}$ = $\nu_{max}/ \Delta\nu$. The parameters $\epsilon$ follows a scaling law $\epsilon = A + Blog\Delta\nu$, and $\alpha(\Delta\nu)$ is related to the large separation $\alpha(\Delta\nu)$=0.015$\Delta\nu^{-0.32}$ \citep{2010A&A...517A..22M,Mosser2011b,Mosser2012a}. The observational frequencies whose values are closest to the predicted pure p-mode frequencies are classified as the p-dominated mixed modes. For theoretical frequencies, we compute the normalized mode inertia \citep{2010aste.book.....A,2017A&ARv..25....1H}:

\begin{equation}
I=\frac{4\pi\int_{0}^{R}[\xi_{r}(r)^{2}+l(l+1)\xi_{h}(r)^{2}]\rho_{0}r^{2}dr}{M[\xi_{r}(R_{phot})^{2}+l(l+1)\xi_{h}(R_{phot})^{2}]},
\label{e3}
\end{equation}

where $\xi_{r}(r)$ and $\xi_{h}(r)$ are the radial and horizontal displacement functions, $\rho_{0}$ is the local density, $R_{phot}$ is the photospheric radius. We define the mixed mode with the lowest mode inertia $I$ per radial order as the p-dominated mixed mode, and the remaining modes within the same radial order as g-dominated mixed modes. 
The observation frequencies for KIC 9145955, KIC 9970396, KIC 9882316, and KIC 11968334 are presented in Table \ref{ta1}, \ref{ta2}, \ref{ta3}, and \ref{ta4}, respectively. 

To measure the agreement between observed and model frequencies, we use goodness-of-fit functions based on a generalized $\chi^{2}$ formalism:

\begin{equation}
\chi^{2}_{all}=\frac{1}{K_{all}}\sum_{i=1}^{K_{all}}\left(\frac{(\nu_{i(all)}^{\rm obs}-\nu_{i(all)}^{\rm mod}}{\sigma^{obs}_{i(all)}}\right)^2,
\label{e4}
\end{equation}

\begin{equation}
\chi^{2}_{l=1g}=\frac{1}{K_{l=1g}}\sum_{i=1}^{K_{l=1g}}\left(\frac{(\nu_{i(l=1g)}^{\rm obs}-\nu_{i(l=1g)}^{\rm mod}}{\sigma^{obs}_{i(l=1g)}}\right)^2,
\label{e5}
\end{equation}
	
where the superscripts “obs” and “mod” refer to observed and model frequencies, respectively. The subscripts ``all" and ``$l = 1g$" indicate whether all modes or only g-dominated mixed modes are included in the sum. $\sigma^{obs}$ represents the observational error for each frequency. $K_{all}$ and $K_{l=1g}$ are the total number of observed frequencies and the number of observed g-dominated mixed modes, respectively. A lower $\chi^{2}$ value indicates a better agreement between the theoretical and observed frequencies. For each evolutionary track, the best-fitting model is the model with the minimum value, e.g., $\chi^{2}_{m,l=1g}$ of $\chi^{2}_{l=1g}$, among all theoretical models.

\section{Results and Discussion} \label{sec4}

\subsection{Best-Fitting Model}\label{s4.1}

\begin{figure*}[t]
\centering
\includegraphics[angle=0,width=\textwidth]{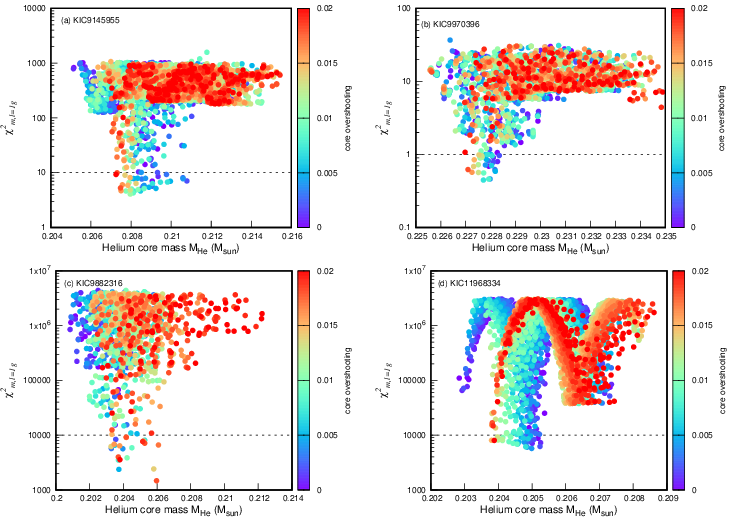}
\caption{(a)$\chi^{2}_{m,l=1g}$ as a function of the helium core mass $M_{He}$ for models of KIC 9145955. (b)$\chi^{2}_{m,l=1g}$ as a function of the helium core mass $M_{He}$ for models of KIC 9970396. (c)$\chi^{2}_{m,l=1g}$ as a function of the helium core mass $M_{He}$ for models of KIC 9882316. (d)$\chi^{2}_{m,l=1g}$ as a function of the helium core mass $M_{He}$ for models of KIC 11968334. Each circle represents a model with the minimal value of $\chi^{2}_{l=1g}$ on each evolutionary track.}
\label{f1}
\end{figure*}

Based on the input physics summarized in Table \ref{t2}, we compute evolutionary model grids for the four red giant stars KIC 9145955, KIC 9970396, KIC 9882316, and KIC 11968334. Figure \ref{f1} shows $\chi^{2}_{m,l=1g}$ as a function of $M_{\rm He}$, where each point corresponds to the model with the minimum $\chi^{2}_{l=1g}$ along a given track. As shown in Figure \ref{f1}, the values of $\chi^{2}_{m,l=1g}$ for KIC 9882316 and KIC 11968334 are extremely large compared to those of KIC 9145955 and KIC 9970396. This difference does not reflect poorer physical modeling for the former two stars, but rather arises directly from the much higher precision of their observational frequency measurements. The frequency tables in the Appendix confirm that the median $l=1g$ uncertainties are $0.036\,\mu$Hz for KIC~9145955 and $0.034\,\mu$Hz for KIC 9970396, in contrast to only $0.0012\,\mu$Hz for KIC 9882316 and $0.0010\,\mu$Hz for KIC 11968334. Since $\chi^{2}$ scales as $\sigma_{\nu}^{-2}$, these ${\sim}30$--35 times smaller uncertainties account for the ${\sim}10^{3}$--$10^{4}$ times larger $\chi^{2}_{l=1g}$ values. Under such conditions, the minimum-$\chi^2$ model at any discrete grid point has a large stochastic component. As noted by \citet{Gruberbauer2012}, when the observational uncertainties are small compared to the frequency differences between adjacent grid points, the $\chi^2$ contrast between models is artificially inflated, and simple $\chi^2$ minimization may not identify the physically most adequate model. To obtain robust grid-based parameter estimates, we compute likelihood-weighted means by marginalizing over the entire model grid. Following the standard approach of evaluating relative likelihoods via 
$\Delta\chi^2 = \chi^2 - \chi^2_{\rm min}$ 
\citep[][Section~15.6]{press2007numerical}, 
we define the rescaled likelihood for model $i$ as

\begin{equation}
\mathcal{L}_i = \exp\left( -\frac{1}{2} \frac{\chi^2_i - \chi^2_{\mathrm{min}}}{\chi^2_{\mathrm{min}}} \right),
\label{e6}
\end{equation}

where $\mathcal{L}$ is the likelihood function, $\chi^{2}_{\min}$ is the minimum $\chi^{2}$ value among all models. The normalization by $\chi^{2}_{\min}$ in the denominator accounts for the case where $\chi^{2}_{\min} \gg 1$, which would otherwise cause all likelihoods to vanish numerically. The likelihood-weighted mean and standard deviation of each fundamental stellar parameter $p$ are then computed as
\begin{equation}
\langle p \rangle = \frac{\sum_{i=1}^{N} \mathcal{L}_i p_i}{\sum_{i=1}^{N} \mathcal{L}_i},
\label{e7}
\end{equation}

\begin{equation}
 \sigma_p  = \left( \frac{\sum_{i=1}^{N} \mathcal{L}_i (p_i - \langle p \rangle)^2}{\sum_{i=1}^{N} \mathcal{L}_i} \right)^{1/2},
\label{e8}
\end{equation}

where $p$ represents any fundamental stellar parameter of interest (e.g., mass, radius, age), $\langle p\rangle$ is the likelihood weighted mean value of the fundamental parameters, $\sigma_{p}$ is standard deviation, and $N$ is the total number of models in the grid. This likelihood-weighted marginalization aligns with the probabilistic framework advocated by \citet{Gruberbauer2012},  who argued that Bayesian marginalization over  model grids provides more robust parameter  estimates than simple $\chi^2$ minimization, particularly when the observational precision exceeds the frequency resolution of the model grid. We use $\chi^{2}_{l=1g}$ to compute the likelihood weights, and the resulting marginalized parameters for all four stars are presented in Table~\ref{t3}. To quantify how well our models physically reproduce the observations independently of the observational uncertainty scale, we also report the unnormalized mean squared residuals $S^{2}_{all}$ and $S^{2}_{l=1g}$ in Table~\ref{t4}, defined as 

\begin{equation}  
S^{2}_{all}=\frac{1}{K_{all}}\sum_{i=1}^{K_{all}}  
\left(\nu_{i(all)}^{\rm obs}-\nu_{i(all)}^{\rm mod}\right)^2,  
\label{e9}  
\end{equation}  
  
\begin{equation}  
S^{2}_{l=1g}=\frac{1}{K_{l=1g}}\sum_{i=1}^{K_{l=1g}}  
\left(\nu_{i(l=1g)}^{\rm obs}-\nu_{i(l=1g)}^{\rm mod}\right)^2,  
\label{e10}  
\end{equation}  

where the symbols are defined as in equations (\ref{e4}) and (\ref{e5}). Unlike $\chi^{2}_{l=1g}$, which weights each squared residual by $\sigma_{\nu,i}^{-2}$, $S^{2}_{l=1g}$ measures only the absolute frequency mismatches and is therefore independent of the adopted uncertainties. For the four best-fitting models, $S^{2}_{l=1g}$ is of the same order of magnitude (Table \ref{t4}), with RMS residuals $\sqrt{S^{2}_{l=1g}}\sim 0.04$--$0.06~\mu$Hz, indicating that the absolute frequency agreement is comparable across all four targets. However, $\chi^{2}_{l=1g}$ differs by several orders of magnitude. This discrepancy can be understood through the effective uncertainty scale $\sigma_{\rm eff}\equiv\sqrt{S^{2}_{l=1g}/\chi^{2}_{l=1g}}$, which is   
$\sim 10^{-2}$--$10^{-1}~\mu$Hz for KIC 9145955 and KIC 9970396 but only $\sim 10^{-3}~\mu$Hz for KIC 9882316 and KIC 11968334, directly reflecting the much smaller reported uncertainties for the latter two stars. The comparable $\sqrt{S^2_{l=1g}}$ values across  all four targets therefore confirm that the elevated $\chi^2_{l=1g}$ for KIC~9882316 and KIC~11968334 reflects their higher measurement precision rather than limitations of the stellar model.

\begin{figure*}[t]
\centering
\includegraphics[angle=0,width=\textwidth]{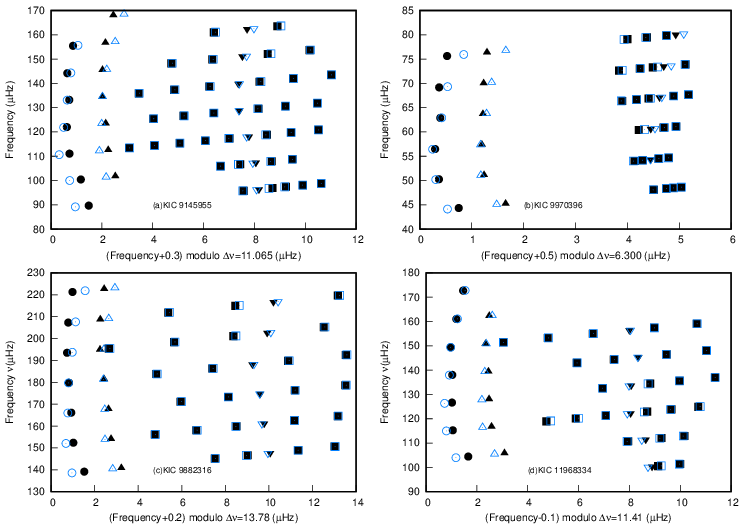}
\caption{Frequency \'{e}chelle diagram of the best-fitting models for KIC 9145955, KIC 9970396, KIC 9882316 and KIC 11968334. The filled and open symbols denote observational and calculated frequencies. The triangles and circles denote $l = 0$ modes and $l = 2$ modes, respectively. The squares and inverted triangles denote g-dominated and p-dominated $l=1$ modes. The fundamental parameters of these four models are presented in Table \ref{t4}.}
\label{f2}
\end{figure*}

In addition to the marginalized parameters in Table~\ref{t3}, we identify a single representative model for each star to serve as the basis for the \'{e}chelle diagrams (Figure~\ref{f2}), propagation diagrams (Figure~\ref{f4}), Kippenhahn diagrams (Figure~\ref{f7}), and the internal structure discussion in subsequent sections. We refer to this as the ``best-fitting model.'' It is selected from the set of asteroseismic candidate models---those with $\chi^{2}_{m,l=1g}$ values at least two orders of magnitude lower than those of surrounding models, as indicated by the horizontal dashed 
lines in Figure~\ref{f1} (threshold values of 10 for KIC~9145955, 1 for  KIC~9970396, and 10{,}000 for both KIC~9882316 and KIC~11968334)---by requiring 
that the model's effective temperature $T_{\rm eff}$, surface gravity $\log g$, and metallicity $[\rm Fe/H]$ are consistent with the spectroscopic observations. 
The model metallicity is calculated from the stellar model's surface hydrogen ($X$) and metal ($Z$) abundances using \citep{2006essp.book.....S}

\begin{equation}
\rm [Fe/H]=lg\left ( \frac{Z}{X}\right )_{star}-lg\left (\frac{Z}{X}\right )_{\odot},
\label{e11}
\end{equation}

where we adopt $(Z/X)_{\odot}$ = 0.0229 \citep{1998SSRv...85..161G}. All reported stellar parameters and their uncertainties are derived from the Bayesian likelihood-weighted marginalization presented in Table~\ref{t3}. The input parameters and derived properties of the best-fitting models are listed in Table~\ref{t4} for reference. $\Delta\nu_{m}$ and $\Delta\Pi_{\rm 1, mod}$ in Table \ref{t3} and \ref{t4} represent model large frequency separation and the theoretical asymptotic period spacing, respectively. We adopt the method from \citep{2022ApJ...927..167L,2023ApJ...953..182W} to obtain the model large separation $\Delta\nu_{m}$:

\begin{equation}
\Delta\nu_{m} = \Delta\nu_{\odot}\left (\frac{\bar{\rho}}{\bar{\rho}_{\odot }} \right )^{0.507},
\label{e12}
\end{equation}
where $\Delta\nu_{\odot}$ = 135.1 $\mu$Hz, $\bar{\rho}_{\odot}$ and $\bar{\rho}$ are the mean solar density and mean stellar density, respectively. The theoretical asymptotic period spacing $\Delta\Pi_{1,mod}$ is defined as \citep{1980ApJS...43..469T,2010aste.book.....A}

\begin{equation}
\Delta\Pi_{1,mod}=\frac{2\pi^{2}}{\sqrt{2}\left (\int_{r1}^{r2}\frac{N}{r}dr \right )},
\label{e13}
\end{equation}
where $r_{1}$ and $r_{2}$ are the inner and outer boundaries of the region where gravity waves propagate, respectively.

The frequency \'{e}chelle diagrams of our best-fitting models (Figure \ref{f2}) show excellent agreement between the theoretical and observed oscillation frequencies, particularly for the g-dominated mixed modes. We also note the presence of systematic offsets for the $l=0$ and $l=2$ modes, which we attribute to the surface effect and discuss further in Section 4.3. The log$g$ values obtained by our best-fitting models are in good agreement with the observations from APOGEE\citep{2022ApJS..259...35A}, LAMOST\citep{2018ApJS..238...30Z}, and \citet{Perez2016}. For metallicity, some small discrepancies are found for specific targets. For KIC 9970396, the model metallicity ($\rm [Fe/H]=-0.128$) is higher than the spectroscopic values from APOGEE ($\rm [Fe/H]=-0.245\pm0.007$; \citealp{2022ApJS..259...35A}) and LAMOST ($\rm [Fe/H]=-0.244\pm0.020$; \citealp{2018ApJS..238...30Z}), though it remains consistent with another APOGEE measurement ($\rm [Fe/H]=-0.18\pm0.07$; \citealp{2015ApJS..219...12A}). Furthermore, its model asymptotic period spacing ($\Delta\Pi_{1,m} = 68.98$ s) is larger than the observed asymptotic period spacing ($\Delta\Pi_{1,obs} = 64\pm0.06$ s), a deviation we analyze in Section 4.2. The other targets show similar but smaller metallicity discrepancies. For KIC 11968334 ($\rm [Fe/H]=0.208$), KIC 9882316 ($\rm [Fe/H]=-0.44$), and KIC 9145955 ($\rm [Fe/H]=-0.31$), the model metallicities are slightly offset from the APOGEE values \citep{2022ApJS..259...35A} but remain consistent with other literature measurements \citep{Perez2016,2018ApJS..238...30Z}. Despite these minor discrepancies, the strong constraints from the high-precision oscillation frequencies support the reliability of these best-fitting models for the target stars.

\begin{figure}[h]
\centering
\includegraphics[angle=0,scale=1.0]{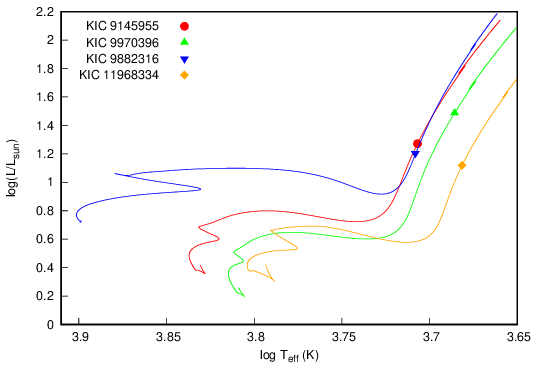}
\caption{The evolution track of KIC 9145955, KIC 9970396, KIC 9882316, and KIC 11968334. The fundamental parameters of these four evolutionary tracks are presented in the Table \ref{t4}. Filled symbols indicate the current position of the best-fitting model.}
\label{f3}
\end{figure}

\begin{figure}[h]
\centering
\includegraphics[angle=0,scale=1.0]{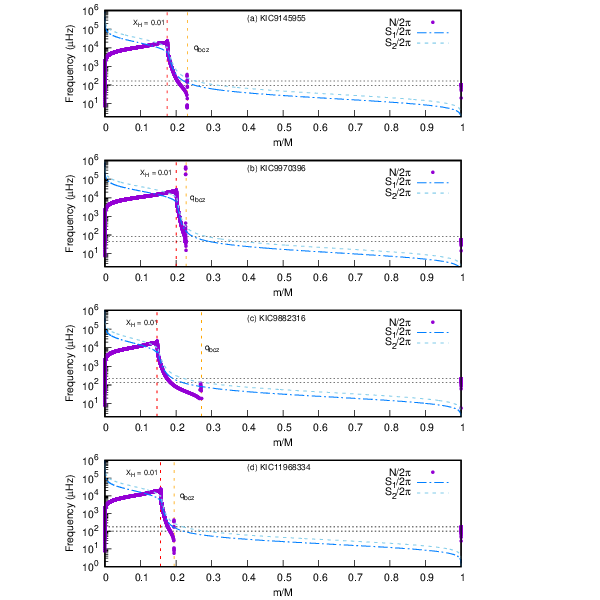}
\caption{Propagation diagram of KIC 9145955, KIC 9970396, KIC 9882316 and KIC 11968334. The vertical red dashed lines in panel (a) to panel (d) denote the helium core boundary where the hydrogen mass fraction is about 0.01. The vertical orange dashed lines in panel (a) to panel (d) denote the base of the convection zone, which is marked as $q_{bcz}$. In panel (a) to panel (d), two horizontal dashed lines represent the range of observed frequencies. The purple dot indicates the Brunt-V$\ddot{\rm a}$is$\ddot{\rm a}$l$\ddot{\rm a}$ frequency $N$ and the blue dash-dotted and light-blue dashed line indicates characteristic acoustic frequency $S_{1}$ and $S_{2}$, respectively. }
\label{f4}
\end{figure}

\begin{figure}[h]
\centering
\includegraphics[angle=0,scale=1.0]{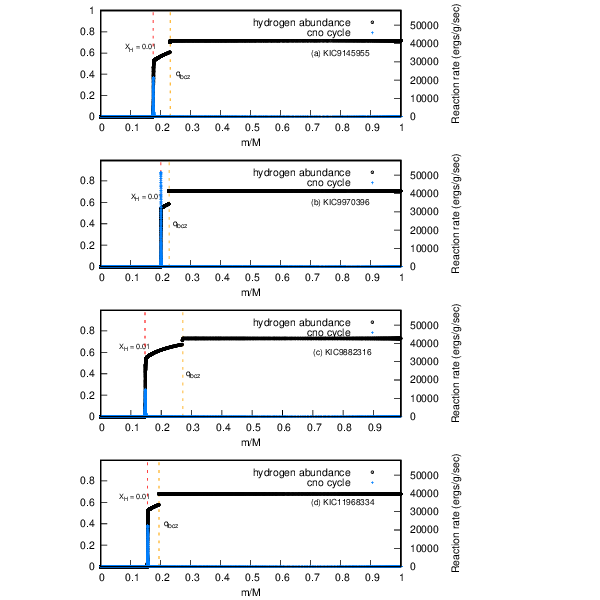}
\caption{Internal structure of KIC 9145955, KIC 9970396, KIC 9882316 and KIC 11968334. The vertical red dashed lines in panel (a) to panel (d) denote the helium core boundary where the hydrogen mass fraction is about 0.01. The vertical orange dashed lines in panel (a) to panel (d) denote the base of the convection zone, which is marked as $q_{bcz}$. The blue crosses in panel (a) to panel (d) represent the reaction rate of the carbon, nitrogen and oxygen (CNO) cycle. }
\label{f5}
\end{figure}

The best-fitting models provide valuable insights into the internal structures and evolutionary states of the four red giant stars. Figure \ref{f3} shows their evolutionary tracks and current locations in the HR diagram, illustrating that the stars occupy different positions along the red giant branch as a consequence of both their masses and evolutionary progression. In particular, although KIC 9970396 ($M=1.14 M_{\odot}$) is not the most massive star in the sample, its location along the track indicates a more advanced evolutionary state compared to the other targets. Figures \ref{f4} and \ref{f5} further reveal pronounced structural differences that correlate with these evolutionary states. For KIC 9970396, the base of the convective zone is located closer to the hydrogen-burning shell in terms of relative mass coordinate, whereas in the more massive star KIC 9882316 ($M=1.40M_{\odot}$), the convective boundary remains farther out in relative mass coordinate. The determination of the base of the convection zone is described in the Appendix. The Brunt-V$\ddot{\rm a}$is$\ddot{\rm a}$l$\ddot{\rm a}$ frequency near the base of the convective envelope exhibits significant variations among the stars, as shown in Figure \ref{f4}, consistent with this structural arrangement. In KIC 9882316, the peak value of the Brunt-V$\ddot{\rm a}$is$\ddot{\rm a}$l$\ddot{\rm a}$ frequency lies below the observed frequency range, while in KIC 9970396 it reaches values of the order of $10^{5}\mu\mathrm{Hz}$. This reflects the more compact core structure and more advanced internal evolution of KIC 9970396 along the red giant branch, given its lower mass. These differences are also manifested in the nuclear energy generation rates. As shown in Figure \ref{f5}, the CNO-cycle energy generation rate in KIC 9970396 reaches approximately $5\times10^{4}\,\mathrm{erg}\,\mathrm{g}^{-1}\,\mathrm{s}^{-1}$, which is significantly higher than in the other stars. These results demonstrate that mixed modes are sensitive probes of internal structure and evolutionary state in red giants, allowing us to distinguish stars at different stages of red giant branch evolution.

\begin{table*}
\caption{\label{t3}Physical Parameters of KIC 9145955, KIC 9970396, KIC 9882316 and KIC 11968334 constrained by the g-dominated mixed modes.}
\centering
\begin{tabular}{lccccc}
\hline
\hline
Parameters                    &KIC9145955 &KIC9970396 &KIC9882316 &KIC11968334 \\
\hline
Mass (M/$M_{\odot}$)                &1.22$\pm$0.03        &1.15$\pm$0.03      &1.40$\pm$0.02      &1.34$\pm$0.02\\
Metal fraction (Z)                  &0.0012$\pm$0.003     &0.012$\pm$0.002    &0.007$\pm$0.001    &0.027$\pm$0.004\\
Overshooting parameter ($f_{ov}$)   &0.008$\pm$0.005      &0.008$\pm$0.005    &0.013$\pm$0.006    &0.006$\pm$0.005\\
Effective temperature ($\rm T_{eff}$/K)  &5009$\pm$60     &4839$\pm$40        &5076$\pm$24        &4765$\pm$29\\
Surface gravity log(g/cgs)               &3.030$\pm$0.003 &2.706$\pm$0.003    &3.172$\pm$0.002    &3.063$\pm$0.002\\
Radius (R/$R_{\odot}$)                   &5.59$\pm$0.04   &7.87$\pm$0.06      &5.08$\pm$0.03      &5.63$\pm$0.03\\
Luminosity ($\rm L_{\odot}$)             &17.70$\pm$0.83  &30.55$\pm$1.10     &15.38$\pm$0.33     &14.69$\pm$0.37\\
Large frequency separation ($\Delta\nu_{m}$/$\mu$Hz)    &10.911$\pm$0.007 &6.280$\pm$0.005 &13.518$\pm$0.010 &11.301$\pm$0.008\\
Period spacing ($\Delta\Pi_{1,mod}$/s)                    &76.98$\pm$0.02   &68.85$\pm$0.36  &80.42$\pm$0.01   &78.08$\pm$0.01\\
Age (t/Gyr)                                             &4.88$\pm$0.45    &6.25$\pm$0.51   &2.67$\pm$0.18    &4.63$\pm$0.31\\
Mass of helium core ($M_{He}$/$M_{\odot}$)   &0.2085$\pm$0.0008   &0.2283$\pm$0.0012  &0.2045$\pm$0.0010  &0.2047$\pm$0.0004\\
Radius of helium core ($R_{He}$/$R_{\odot}$) &0.03068$\pm$0.00009 &0.03007$\pm$0.00012	 &0.03091$\pm$0.00011 &0.03132$\pm$0.00014\\
\hline
\end{tabular}
\end{table*}

\begin{table*}
\caption{\label{t4}Input and Fundamental Parameters of the best-fitting models.}
\centering
\begin{tabular}{lcccc}
\hline
\hline
Parameters                            &KIC9145955 &KIC9970396 &KIC9882316 &KIC11968334 \\
\hline
&&Input Parameters\\
\hline
Mass (M/$M_{\odot}$)                  &1.20  &1.14  &1.40  &1.31\\
Initial Metal fraction (Z)            &0.008 &0.011 &0.006 &0.024\\
Overshooting parameter ($f_{ov}$)     &0.007 &0.005 &0.015 &0.009 \\
Mixing length parameter ($\alpha$)    &2.0   &2.0   &2.0   &2.0\\
\hline
&&$\chi^{2}$ Value\\
\hline
$\chi_{l=1g}^{2}$                                &6.730   &0.588    &7816.585 &6889.805\\
$\chi_{all}^{2}$                                 &131.731 &11.082   &4450.201 &6499.917\\
$S^{2}_{l=1g}$                                   &0.00250 &0.00170  &0.00363  &0.00383\\
$S^{2}_{all}$                                    &0.0280 &0.0129    &0.0420   &0.0225\\
\hline
&&Fundamental Parameters\\
\hline
Effective temperature ($\rm T_{eff}$/K)  &5093  &4851  &5106  &4782 \\
Surface gravity log (g/cgs)              &3.027 &2.705 &3.171 &3.060\\
Radius (R/$R_{\odot}$)                   &5.56  &7.85  &5.09  &5.59 \\
Luminosity ($\rm L_{\odot}$)             &18.67  &30.67  &15.79  &14.7\\
Large frequency separation ($\Delta\nu_{m}$/$\mu$Hz) &10.909 &6.283 &13.500 &11.302\\
Period spacing ($\Delta\Pi_{1,mod}$/s)               &76.96  &68.98 &80.39  &78.07\\
Age (t/Gyr)                                      &4.63   &6.27  &2.51   &4.85 \\
Mass of helium core ($M_{He}$/$M_{\odot}$)       &0.2095  &0.2282   &0.2052  &0.2046 \\
Radius of helium core ($R_{He}$/$R_{\odot}$)     &0.03062 &0.03002  &0.03085 &0.03113\\
\hline
\end{tabular}
\end{table*}

\subsection{The Period Spacing of KIC 9970396}\label{s4.2}

KIC 9970396 is a pulsating red giant in a detached eclipsing binary system, first identified by \citet{2013ApJ...767...82G, 2014ApJ...785....5G}. The evolutionary stage of KIC 9970396 is between the first dredge-up and the red giant bump phase \citep{2020MNRAS.494..511Z}. The mass and radius of KIC 9970396 are constrained by dynamical binary
measurements, with reported mass and radius values of: $M=1.14\pm0.03M_{\odot}$, $R=8.0\pm0.2R_{\odot}$ \citep{2016ApJ...832..121G}, and $M=1.178\pm0.015M_{\odot}$, $R=8.035\pm0.074M_{\odot}$ \citep{2018MNRAS.476.3729B}. The mixing-length parameter $\alpha$ has been determined as $2.2\pm0.23$ \citep{2018MNRAS.475..981L} and 2.0 \citep{2020MNRAS.494..511Z} based on asteroseismic modeling. In this study, we derive a mass of $1.15\pm0.02M_{\odot}$ and a radius of $7.88\pm0.05R_{\odot}$. These parameters are in excellent agreement with the dynamical solutions, with our radius being lower than that of \citet{2018MNRAS.476.3729B} by only 0.4$\%$. This consistency provides strong support for the accuracy and reliability of our stellar models.

We employ the method of \citet{2024A&A...688A.184L} to characterize the mixed-mode pattern of KIC 9970396. This method accounts for rotational perturbations and solves the mixed-mode coupling equations, allowing for the simultaneous derivation of core and envelope rotation rates as well as the asymptotic period spacing. Using this approach, we derived an observed asymptotic period spacing of $\Delta\Pi_{1,\rm obs} = 64 \pm 0.06$ s. The best-fitting model, however, predicts a theoretical asymptotic spacing of $\Delta\Pi_{1,\rm model} = 68.98$ s (calculated via the integral of the Brunt-V$\ddot{\rm a}$is$\ddot{\rm a}$l$\ddot{\rm a}$ frequency, Eq. \ref{e10}). While there is a difference between the derived observational parameter and the theoretical integral value, the individual mode frequencies provide a more stringent test of the internal structure. As shown in the \'{e}chelle diagram in Figure \ref{f2}(b), the calculated frequencies of the g-dominated mixed modes (open symbols) show excellent agreement with the observational data (filled symbols). This indicates that our stellar model correctly reproduces the internal structure of KIC 9970396. The apparent offset between $\Delta\Pi_{1,\rm obs}$ and $\Delta\Pi_{1,\rm model}$ likely arises from the sensitivity of the extraction method to specific mode trapping or coupling effects, rather than a fundamental deficiency in the stellar model. Consequently, we rely on the robust agreement of individual mixed-mode frequencies to validate our analysis, confirming that the model accurately captures the evolutionary state of the star.

\subsection{Surface Effect}\label{s4.3}

The surface effect is the systematic difference between calculated and observed oscillation frequencies caused by imperfect modelling of a star's surface in physical structures \citep{1988Natur.336..634C,1996Sci...272.1286C,1997MNRAS.284..527C}. This effect primarily impacts pure p modes, reducing the agreement between models and observations\citep{2008ApJ...683L.175K}. On the other hand, when it comes to mixed modes, and more specifically, those g-dominated mixed modes, the impact of the surface effect is less evident \citep{2017A&ARv..25....1H}.

In this study, we use the correction method proposed by \citet{2014A&A...568A.123B} to correct the surface effect in oscillation frequencies of the best-fitting model of KIC 9145955, KIC 9970396, KIC 9882316, and KIC 11968334. 
The correction formula combines inverse and cubic frequency terms, scaled by the inverse of the normalized mode inertia:

\begin{equation}
\delta\nu=\left (a_{-1}(\nu/\nu_{ac})^{-1}+a_{3}(\nu/\nu_{ac})^3 \right )/I,
\label{e14}
\end{equation}

where $a_{-1}$ and $a_{3}$ are fitting coefficients, the method to determine $a_{-1}$ and $a_{3}$ was described by \citet{2014A&A...568A.123B}. $I$ is the normalized mode inertia which has been defined in equation (\ref{e2}). $\nu_{ac}$ is the acoustic cutoff frequency and is defined by
\begin{equation}
\nu_{ac}/\nu_{ac,\odot}=\frac{g}{g_{\odot}}\left ( \frac{T_{\rm eff}}{T_{\rm eff,\odot}}\right )^{-1/2},
\label{e15}
\end{equation} 

\begin{figure*}[t]
\centering
\includegraphics[angle=0,width=\textwidth]{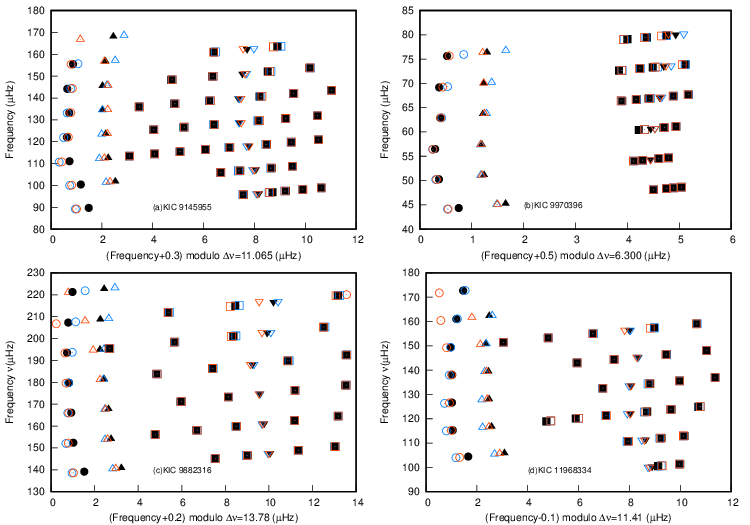}
\caption{Frequency \'{e}chelle diagrams before and after applying the surface effect correction of KIC 9145955, KIC 9970396, KIC 9882316, and KIC 11968334. The filled and open symbols denote observational and calculated frequencies. The triangles and circles denote $l = 0$ modes and $l = 2$ modes, respectively. The squares and inverted triangles denote the g-dominated and p-dominated $l = 1$ modes. Blue symbols represent the theoretical frequencies before the surface effect correction is applied. Red symbols represent the theoretical frequencies after surface effect correction is applied. The fundamental parameters of these models are presented in Table \ref{t4}.}
\label{f6}
\end{figure*}

We adopt $\nu_{ac,\odot}$ = 5000 $\mu$Hz \citep{2014A&A...568A.123B}, $\log g_{\odot}$ = 4.44, and $T_{\rm eff,\odot}$ = 5777 K \citep{2000asqu.book.....C}. Figure \ref{f6} presents the \'{e}chelle diagrams of KIC 9145955, KIC 9970396, KIC 9882316,and KIC 11968334 after applying the surface correction to the frequencies. 

As shown in Figure \ref{f6}, the application of surface corrections results in minimal changes to the g-dominated mixed modes across all four red giants. This stands in stark contrast to the pure acoustic ($l=0$) and p-dominated modes, which are significantly affected by the corrections. While the applied surface correction remains imperfect for p-modes, particularly at higher frequencies, the relative insensitivity of g-dominated mixed modes highlights their robustness against surface uncertainties. Consistent with our previous findings \citep{2018ApJ...855...16Z,2020MNRAS.494..511Z}, the frequency offsets for these modes are negligible. Consequently, we conclude that surface-effect corrections are unnecessary when employing $\chi_{l=1g}^2$ to constrain stellar models.

\subsection{Convective Core Overshooting}\label{s4.4}

\begin{figure*}[t]
\centering
\includegraphics[angle=0,width=\textwidth]{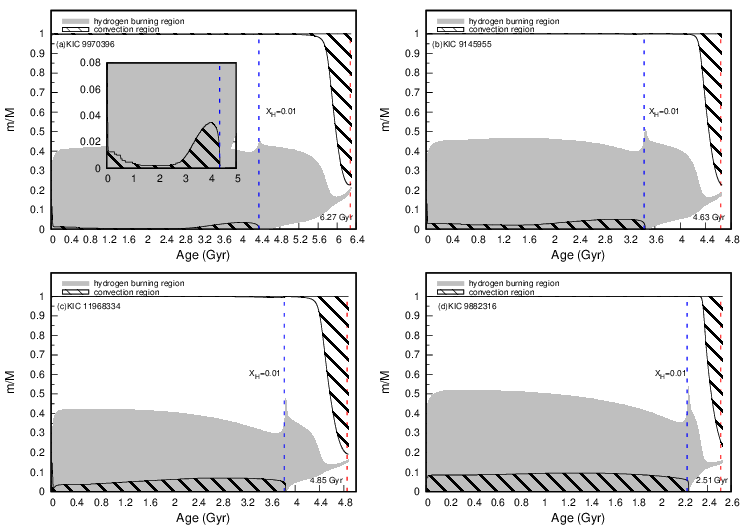}
\caption{Kippenhahn diagrams of the best-fitting models for KIC 9145955, KIC 9970396, KIC 9882316 and KIC 11968334. Regions where convection takes place are hatched. Regions where nuclear burning produces more than 1 erg/(g $\times$ s) are shown in grey for hydrogen burning. The vertical blue dashed lines in both panels indicate the age at the end of the main-sequence stage, that is, when the core hydrogen mass fraction is about 0.01. The vertical red dashed lines in both panels indicate the current age of KIC 9145955, KIC 9970396, KIC 9882316 and KIC 11968334, respectively.}
\label{f7}
\end{figure*}

Based on our analysis of four red giants spanning a mass range from 1.14 to 1.40 $M_{\odot}$, we investigate the potential mass dependence of the core overshooting parameter ($f_{ov}$). Table \ref{t4} presents the best-fitting overshooting parameters, and Figure \ref{f7} presents the corresponding Kippenhahn diagrams ordered by increasing stellar mass, visualizing the evolution of convective regions during the main sequence. The lowest-mass star in our sample, KIC 9970396 ($1.14 M_{\odot}$), exhibits only a small convective core. Its best-fitting asteroseismic model provides a low overshooting parameter ($f_{ov}=0.005$). The intermediate-mass stars KIC 9145955 ($1.20 M_{\odot}$) and KIC 11968334 ($1.31 M_{\odot}$) have larger convective cores, resulting in intermediate overshooting values of $f_{ov}=0.007$ and $f_{ov}=0.009$, respectively. The most massive star in our sample, KIC 9882316 ($1.40 M_{\odot}$), develops a substantially larger convective core extending to approximately 8\% of the stellar mass during the main-sequence phase (Figure \ref{f7}), and its best-fitting model requires a notably larger overshooting parameter of $f_{ov}=0.015$.  
When the four targets are ordered by increasing stellar mass (1.14,1.20,1.31,1.40), their corresponding convective core overshooting $f_{ov}$ values (0.005, 0.007, 0.009, 0.015) also form a monotonically increasing sequence. While this result does not constitute a statistically conclusive detection given the small sample size, our findings are consistent with several prior studies that have investigated the mass dependence of convective overshooting. For instance, \citet{2010ApJ...718.1378M} proposed a mass-dependent relation $f_{ov} = (0.13~M/M_{\odot} - 0.098)/9$ for stars in the mass range $1.1$--$2.0~M_{\odot}$, with a fixed value of $f_{ov} = 0.018$ adopted above $2.0~M_{\odot}$. \citet{2016A&A...592A..15C,2017ApJ...849...18C,2018ApJ...859..100C,2019ApJ...876..134C} also provided empirical evidence from the study of 50 well-studied double-lined eclipsing binaries that $f_{ov}$ rises sharply with stellar masses in the range of 1.2 to 2 $M_{\odot}$, while $f_{ov}$ remains constant in the range from 2 $M_{\odot}$ up to 4.4 $M_{\odot}$. Similar increasing trends for low-mass stars have also been reported by \citet{2019ApJ...879...86G} and \citet{2022MNRAS.512.4852Z}. Although a definitive confirmation of this trend requires a substantially larger asteroseismic sample, our detailed modeling of individual oscillation frequencies provides precise, independent constraints on $f_{ov}$ for individual red giants, reinforcing the conclusion that asteroseismology provides valuable and independent constraints on the convective core overshooting parameter.

\subsection{The Mixing-Length Parameter}\label{s4.5}

In our previous work \citep{2018ApJ...855...16Z,2020MNRAS.494..511Z}, we found that the size of the helium core is less affected by the different mixing-length parameters. The $\alpha$ for KIC 9145955 and KIC 9970396 is determined to be 2.0 with the Eddington gray atmosphere as the boundary conditions. For KIC 9882316 and KIC 11968334, we adopt a simple photosphere atmospheric model as the boundary condition in our stellar models. In a relevant study, \citet{2018MNRAS.475..981L} utilized models with a simple photospheric boundary condition and determined the mixing-length parameter for red giants to be $1.14 \pm 0.07$ times the calibrated solar value. Although we adopt $\alpha=2.0$ for our stellar models, we investigate the impact of varying the mixing-length parameter on the fundamental properties of KIC 9882316, which uses the photosphere atmospheric model. This star serves as an appropriate test case because its metallicity measurements are consistent across different methods within observational uncertainties. The initial parameters for this investigation, which include convective core overshooting, are listed in Grid B of Table \ref{t2}. For low-mass red giants, the mixing-length parameter $\alpha$ primarily influences the stellar radius and the structure of the envelope. Because p modes are highly sensitive to the envelope, we employ all available oscillation frequencies, including g-dominated mixed modes, p-dominated mixed modes, and $l=0$ and $l=2$ modes, as observational constraints. We then compare the resulting stellar parameters with those obtained using only the g-dominated mixed modes.

\begin{figure*}[t]
\centering
\includegraphics[angle=0,width=\textwidth]{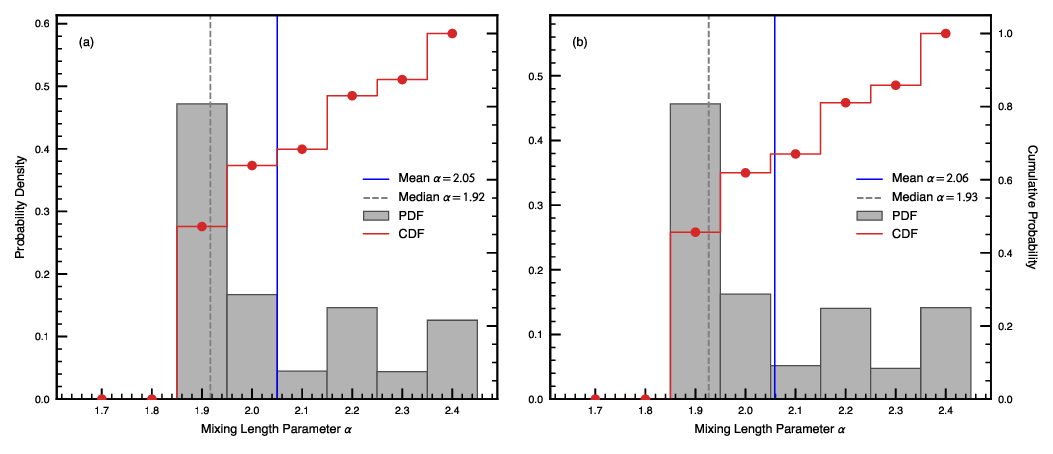}
\caption{Panel (a) The cumulative distribution of $\chi^2_{l=1g}$ values as a function of the mixing-length parameter $\alpha$. Panel (b) The cumulative distribution of $\chi^2_{all}$ values as a function of the mixing-length parameter $\alpha$. The blue vertical lines indicate the probability-weighted mean values derived from the Bayesian marginal likelihood analysis, and the dashed line represents the median value. The red stepped line with filled circles represents the Cumulative Distribution Function (CDF). The grey bars represent the Probability Density Function. }
\label{f8}
\end{figure*}

Based on the results in Section \ref{s4.4}, red giants with masses greater than 1.4 $M_\odot$ generally require $f_{ov}>0.01$. Therefore, for KIC 9882316, we restrict the convective core overshooting parameter $f_{ov}$ in Grid B to the range 0.010-0.015. To robustly determine the mixing-length parameter, we computed the probability-weighted mean of $\alpha$ across the model grid using a Bayesian marginal likelihood approach. To conservatively account for model systematics given the high precision of the asteroseismic data, we rescaled the likelihood such that the effective degrees of freedom match the global minimum $\chi^2$. Figure \ref{f8} displays the Probability Density Function (PDF) and Cumulative Distribution Function (CDF) of $\alpha$ derived from this analysis for g-dominated mixed modes (panel a) and all observational constraints (panel b). As illustrated in Figure \ref{f8}, the PDF (grey bars) reveals a distribution with the dominant peak at $\alpha \approx$ 1.9 and a heavy tail towards higher $\alpha$ values. This multi-modal structure, notably the secondary probability masses at $\alpha = 2.2$ and 2.4, is a common manifestation of parameter degeneracy in stellar modeling. Specific combinations of other fundamental parameters (such as mass and metallicity) can partially compensate for the structural changes induced by a higher mixing-length parameter, leading to local likelihood maxima. While the median (grey dashed line) is around 1.92, the significant probability mass at $\alpha = 2.2$ and 2.4 shifts the expectation value (mean, blue solid line) to higher values. We adopt this Bayesian expectation value as it robustly accounts for these inherent parameter degeneracies and incorporates the full range of solutions allowed by the data. Specifically, our analysis yields $\alpha = 2.05 \pm 0.18$ using g-dominated mixed modes and $\alpha = 2.06 \pm 0.18$ using all modes. Given the large uncertainty ($\pm 0.18$), these values are statistically consistent with our adopted grid value of $\alpha=2.0$. The resulting fundamental parameters are listed in Table \ref{t5}. Notably, our derived mixing-length parameter is larger than the solar value ($\alpha_{\odot} \approx 1.86$) for our input physics. This result aligns with previous studies; for instance, \citet{2018MNRAS.475..981L} demonstrated that the mixing-length parameter for red giants is typically $\sim 14\%$ larger than the solar value. Our result is consistent with this trend within the uncertainties. Consequently, we conclude that adopting a value of $\alpha = 2.0$ for our red giant models is physically justified.

\begin{table*}
\caption{\label{t5} Physical Parameters of KIC 9882316 with different mixing-length parameters.}
\centering
\begin{tabular}{lcc}
\hline
\hline
Parameters                    		&$\chi^{2}_{l=1g}$ 	    &$\chi^{2}_{all}$ \\
\hline
Mass(M/$M_{\odot}$)          		&1.39$\pm$0.02		    &1.39$\pm$0.02\\
Metal fraction(Z)            		&0.006$\pm$0.002		&0.006$\pm$0.002\\
Overshooting parameter($f_{ov}$)  	&0.012$\pm$0.002		&0.012$\pm$0.002\\
MLT parameter($\alpha$)  	        &2.05$\pm$0.18			&2.06$\pm$0.18\\
Effective temperature($\rm T_{eff}$/K) 		&5111$\pm$109		&5115$\pm$111\\
Surface gravity log(g/cgs)    				&3.171$\pm$ 0.002	&3.171$\pm$0.002\\
Radius(R/$R_{\odot}$)        				&5.07$\pm$0.03		&5.07$\pm$0.03\\
Luminosity($\rm L_{\odot}$)	    			&15.81$\pm$1.39		&15.85$\pm$1.43\\
Large frequency separation($\Delta\nu_{m}$/$\mu$Hz) &13.519$\pm$0.012		&13.518$\pm$0.013\\
Period spacing($\Delta\Pi_{1}$/s)    				&80.41$\pm$0.02	&80.41$\pm$0.02\\
Age(t/Gyr)                           				&2.58$\pm$0.28	&2.58$\pm$0.29\\
Mass of helium core($M_{He}$/$M_{\odot}$)   	&0.2048$\pm$0.0023		&0.2048$\pm$0.0024\\
Radius of helium core($R_{He}$/$R_{\odot}$) 	&0.03084$\pm$0.00012		&0.03085$\pm$0.00013\\
\hline
\end{tabular}
\end{table*}

\subsection{Rotation Splitting}\label{s4.6}

\begin{figure*}[t]
\centering
\includegraphics[angle=0,width=\textwidth]{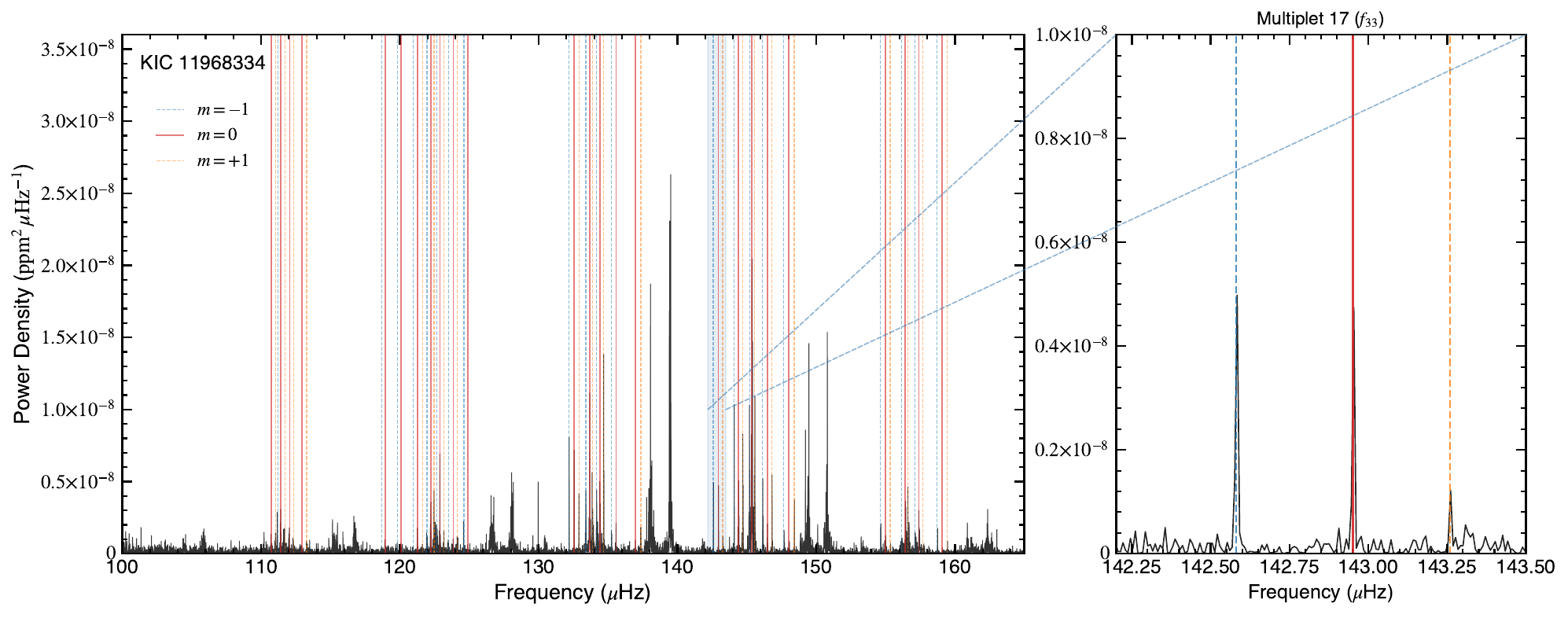}
\caption{Power density spectrum of KIC~11968334 computed from the \textit{Kepler} long-cadence photometry. The $l=1$ oscillation frequencies listed in Table~\ref{t6}, as identified by \citet{corsaro2015}, are indicated by vertical lines: blue dashed lines, red solid lines, and orange dashed lines correspond to the $m=-1$, $m=0$, and $m=+1$ components of each rotational multiplet, respectively. The shaded region indicates the frequency range shown in the right panel. The right panel shows a zoom-in view of Multiplet~17 ($f_{33}$; $142.2$--$143.5\,\mu$Hz), illustrating a clearly resolved rotational triplet.}
\label{f9}
\end{figure*}

A stellar pulsation mode is characterized by three quantum numbers: radial order $n$, spherical degree $l$, and azimuthal order $m$ \citep{2010aste.book.....A}. In a spherically symmetric star, modes with the same $n$ and $l$ but different $m$ values are degenerate, meaning they share the same frequency. Stellar rotation breaks this symmetry, lifting the degeneracy and thereby splitting a single non-radial oscillation frequency into $2l+1$ distinct components. The rotational splittings measured for our target, KIC 11968334, are presented in Table \ref{t6} and form the basis of the following analysis.
The rotational splittings measured for our target, KIC 11968334, as originally extracted by \citet{corsaro2015}, are presented in Table \ref{t6} and visually illustrated in the power density spectrum in Figure \ref{f9}.
For evolved stars, rotation is usually slow enough that second-order effects such as centrifugal distortion can be neglected. The first-order rotational splitting can be expressed as \citep{2010aste.book.....A}
\begin{equation}
\nu_{nlm} = \nu_{nl0} + m\delta\nu_{nl},
\label{e16}
\end{equation}

where $\delta\nu_{n,l}$ is the rotational splitting. \citet{2013A&A...549A..75G} showed that rotational splitting $\delta\nu_{nl}$ is linearly related to the trapping parameter $\zeta$ for each mode, with coefficients depending on the mean rotation rates in the g-mode and p-mode cavities, $\Omega_{core}$ and $\Omega_{env}$:

\begin{equation}
\delta\nu_{n,l}=\zeta \left ( \frac{ \Omega_{core}}{2}  - \Omega_{env} \right )+ \Omega_{env},
\label{e17}
\end{equation}

where the trapping parameter $\zeta$ indicates how strongly a mode is trapped in the radiative core and is defined as the ratio of mode inertia within the g-mode cavity $I_{g}$ to the total mode inertia $I$:

\begin{equation}
\zeta\equiv \frac{I_{g}}{I} =\frac{\int_{r1}^{r2}[\xi_{r}^{2}+l(l+1)\xi_{h}^{2}]\rho r^{2}dr}{\int_{0}^{R}(\xi_{r}^{2}+l(l+1)\xi_{h}^{2})\rho r^{2}dr},
\label{e18}
\end{equation}

where $r_{1}$ and $r_{2}$ denote the boundaries of the g-mode cavity, $\xi_{r}$ and $\xi_{h}$ are the radial and horizontal displacement eigenfunctions, $\rho$ is the density, and $R$ is the stellar radius. For an $l=1$ mixed mode, modes with $\zeta$ close to 1 are g-dominated mixed modes, meaning the g character dictates the mode's behavior. Conversely, modes with $\zeta$ near 0.5 correspond to p-dominated mixed modes \citep{Mosser2012a}.

\begin{figure}[h]
\centering
\includegraphics[angle=0,scale=1.0]{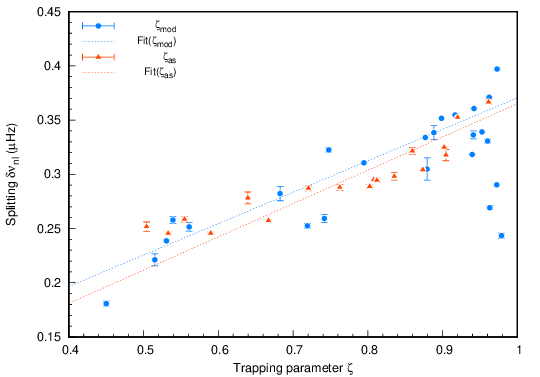}
\caption{Rotational splittings $\delta\nu_{n,l}$ as a function of the mode inertia ratio $\zeta$ for KIC 11968334. The blue open circles represent the observed splittings plotted against the asymptotic $\zeta_{\rm as}$ (derived from the period spacing $\Delta P$), while the red triangles show the same splittings plotted against the model-predicted $\zeta_{\rm mod}$ (derived from our best-fit MESA model). The solid red line and dashed blue line indicate the weighted linear fits for the model and asymptotic values, respectively. The resulting core rotation rates ($\Omega_{\rm core}$) are listed in the legend, showing excellent agreement between the two methods.}
\label{f10}
\end{figure}

In principle, evaluating $\zeta$ from equation (\ref{e18}) requires the displacement eigenfunctions $\xi_{r,h}(r)$, which are only available after constructing detailed asteroseismic models. A key objective of this study is to construct asteroseismic models that accurately reproduce the observed oscillation frequencies, particularly the g-dominated mixed modes. From our best-fitting model of KIC 11968334, we compute the trapping parameter for each mode, denoted as $\zeta_{mod}$. 
An important consideration in this analysis is that the p-mode component of the mixed-mode eigenfunction can be affected by near-surface structural uncertainties (i.e., the surface effect; \citealt{Ong2024}), which may introduce a systematic bias into the envelope rotation rate derived using $\zeta_{\rm mod}$. To assess the robustness of our results against this potential systematic bias, we independently compute the trapping parameter directly from the observed frequencies. Several approaches exist for deriving an observation-based trapping parameter, including analytical methods based on fitted mixed-mode coupling functions \citep{OngGehan2023}. For simplicity, we adopt the widely used asymptotic relation \citep{2013A&A...549A..75G,2014A&A...564A..27D,2015A&A...584A..50M}:
\begin{equation}
\zeta_{\text{as}} = \frac{\Delta P_{\rm obs}}{\Delta \Pi_{\rm 1,obs}}
\label{e19}  
\end{equation}
where $ \Delta P_{\rm obs}$ is the observed bumped period spacing between two consecutive dipole mixed modes and $\Delta \Pi_{\rm 1,obs}$ is the asymptotic period spacing determined from observations. Unlike $\zeta_{\rm mod}$, this quantity is derived entirely from observed frequencies and is therefore free from the bias introduced by surface effects in the model p-mode eigenfunctions. By performing weighted linear fits to the observed rotational splittings $\delta\nu_{n,l}$ as a function of both $\zeta_{\rm mod}$ and $\zeta_{\rm as}$ (Figure \ref{f10}), we independently determine the rotation rates for KIC 11968334. 

As shown in Figure \ref{f10} and Table \ref{t7}, the core rotation rates derived from the two methods are consistent within $1\sigma$: $\Omega_{\rm core} = 0.7409 \pm 0.0113\,\mu$Hz ($\zeta_{\rm mod}$) and $\Omega_{\rm core} = 0.7300 \pm 0.0110\,\mu$Hz ($\zeta_{\rm as}$), a difference of only $\sim$1.5\%. In contrast, the envelope rotation rates exhibit a systematic offset between the two methods: the model-based approach yields $\Omega_{\rm env} = 0.0807 \pm 0.0285\,\mu$Hz, while the asymptotic method gives $\Omega_{\rm env} = 0.0580 \pm 0.0364\,\mu$Hz. Although these two values are consistent within $1\sigma$ given the large uncertainties, the offset consistently shows $\zeta_{\rm mod}$ yielding a higher envelope rotation rate than $\zeta_{\rm as}$, in qualitative agreement with the surface-effect bias predicted by \citet{Ong2024}.

Our derived rotation rates for KIC~11968334 are consistent with the range reported for red giant stars. \citet{2019ARA&A..57...35A} compiled differential rotation profiles for 45 giants and found mean core rotation rates in the range 0.5--1.0~$\mu$Hz and envelope rates between 0.015 and 0.1~$\mu$Hz \citep[see][Figure 4]{2019ARA&A..57...35A}. 
\citet{Triana2017} previously analyzed KIC~11968334 using the same linear relation between rotational splitting and $\zeta$, computing $\zeta$ from both asteroseismic models ($\zeta_{\rm mod}$) and the asymptotic approximation ($\zeta_{\rm as}$).
A detailed comparison of our derived rotation rates and stellar parameters with those of \citet{Triana2017} is presented in Table~\ref{t7}. Our $\zeta_{\rm mod}$-based core and envelope rotation rates of $\Omega_{\rm core} = 0.7409 \pm 0.0113\,\mu$Hz and $\Omega_{\rm env} = 0.0807 \pm 0.0285\,\mu$Hz are consistent within the uncertainties with the corresponding model-based values reported by \citet{Triana2017}: $\Omega_{\rm core} = 0.7295 \pm 0.0222\,\mu$Hz and $\Omega_{\rm env} = 0.0742 \pm 0.0281\,\mu$Hz, respectively. Notably, a systematic offset in the same direction is also present in the results of \citet{Triana2017}: their $\zeta_{mod}$-based envelope rotation rate ($0.0742\mu$ Hz) exceeds the $\zeta_{as}$-based value ($0.0490\mu$Hz) by a comparable margin. The fact that this systematic pattern appears independently in two separate analyses of the same star further supports the interpretation that the offset originates from the surface-effect bias described by \citet{Ong2024}. Some differences nonetheless exist in the underlying stellar models between this work and \citet{Triana2017}, including the derived mass ($1.31\,M_{\odot}$ versus $1.11\,M_{\odot}$), the convective core overshooting parameter ($f_{\rm ov} = 0.009$ versus $f_{\rm ov} = 0.0285$), and the initial metallicity. Specifically, our model adopts an initial metallicity of $Z_{\rm ini} = 0.024$, whereas \citet{Triana2017} assumed ${\rm [Fe/H]}_{\rm ini} = -0.1$. The spectroscopic measurements of ${\rm [Fe/H]}$ compiled in Table~\ref{t1} consistently indicate   
that KIC~11968334 is a metal-rich star; therefore, the metallicity adopted in our model is more consistent with the observational data.   

The discrepancy in the envelope rotation rate between the two methods is consistent with the surface-effect bias mechanism described by \citet{Ong2024}, wherein near-surface structural uncertainties systematically affect the p-mode eigenfunction amplitudes and hence the inferred $\Omega_{env}$ when using $\zeta_{\rm mod}$. The observation-based $\zeta_{\rm as}$ approach circumvents this issue by relying solely on measured frequencies \citep{OngGehan2023}. In addition, the envelope rate is inherently more difficult to constrain because the dipole-mode splittings in low-luminosity giants are primarily sensitive to core rotation \citep{2012ApJ...756...19D}. \citet{2021RvMP...93a5001A} similarly emphasized that near-core rotation can be reliably measured in stars with convective envelopes and radiative cores, the surface rotation rate remains poorly   
determined from mixed-mode splittings alone.

The excellent agreement between our modeled g-dominated mixed-mode frequencies and the observations is of particular importance. As highlighted by \citet{2025A&A...693A.274A}, accurately reproducing mixed-mode properties is essential for obtaining reliable internal rotation profiles. This close model-to-observation agreement validates our calibrated overshooting parameter and supports the reliability of our core rotation measurement. In summary, these results demonstrate that the asteroseismic analysis of mixed modes provides robust constraints on the internal rotation profiles of red giant stars, complementing detailed stellar modeling.

\begin{longtable}{
    @{\extracolsep{3pt}}  
    cccccl@{\quad}cccccc  
}
\caption{Rotation Splitting of KIC 11968334 \label{t6}} \\
\toprule
\toprule
Multiplet & ID & Freq ($\mu$Hz) & $\delta\nu$ ($\mu$Hz) & l & m & 
Multiplet & ID & Freq ($\mu$Hz) & $\delta\nu$ ($\mu$Hz) & l & m \\
\midrule
\endfirsthead

\multicolumn{12}{c}{continued Table \ref{t6}} \\
\toprule
\toprule
Multiplet & ID & Freq ($\mu$Hz) & $\delta\nu$ ($\mu$Hz) & l & m & 
Multiplet & ID & Freq ($\mu$Hz) & $\delta\nu$ ($\mu$Hz) & l & m \\
\midrule
\endhead

\midrule
\multicolumn{12}{r}{continued...} \\
\endfoot
\bottomrule
\endlastfoot

			&$f_{14}$	&110.7372	& 				&1	&0	&			&			&134.1877	&	&1	&-1\\
1	        &			&			&0.3225			&	&	&			&			&			&0.238	&&\\
			&			&111.0597	&				&1	&+1	&14			&$f_{30}$	&134.4257	&	&1	&0\\
			&			&			&				&	&	&			&			&			&0.267	&&\\
			&			&			&				&	&	&			&			&134.6927	&	&1	&+1\\
\\ \hspace*{\fill} \\			
			&			&111.2019	&				&1	&-1	&			&			&135.2455	&	&1	&-1\\
			&			&			&0.2297			&	&	&15			&			&			&0.3363	&&\\
2			&$f_{15}$	&111.4316	&				&1	&0	&			&$f_{31}$	&135.5818	&	&1	&0\\
			&			&			&0.2733			&	&\\
			& 			&111.7049	&				&1	&+1\\      
\\ \hspace*{\fill} \\			
			&$f_{16}$	&112.0552	&				&1	&0	&			&$f_{32}$	&136.9763	&	&1	&0\\
3			&			&			&0.3048			&	&	&16			&			&			&0.397	&&\\
			&			&112.3600	&				&1	&+1	&			&			&137.3733	&	&1	&+1\\
\\ \hspace*{\fill} \\			
			&$f_{17}$	&112.9492	&				&1	&0	&			&			&142.5812	&	&1	&-1\\
4			&			&			&0.3306			&	&	&			&			&			&0.3701	&&\\
			&			&113.2798	&				&1	&+1	&17			&$f_{33}$	&142.9513	&	&1	&0\\
			&			&			&				&	&	&			&			&			&0.3081	&&\\
			&			&			&				&	&	&			&			&143.2594	&	&1	&+1\\
\\ \hspace*{\fill} \\			
			&			&118.6890	&				&1	&-1	&			&			&144.0835	&	&1	&-1\\
5			&			&			&0.2434			&	&	&			&			&			&0.3222	&&\\
			&$f_{20}$	&118.9324	&				&1	&0	&18			&$f_{34}$	&144.4057	&	&1	&0\\
			&			&			&				&	&	&			&			&			&0.299	&&\\
			&			&			&				&	&	&			&			&144.7047	&	&1	&+1\\
\\ \hspace*{\fill} \\			
			&			&119.8195	&				&1	&-1	&			&			&145.1983	&	&1	&-1\\	
6			&			&			&0.2692			&	&	&			&			&			&0.1746	&&\\
			&			&120.0887	&				&1	&0	&19			&$f_{35}$	&145.3729	&	&1	&0\\	 
			&			&			&				&	&	&			&			&			&0.1869	&&\\
			&			&			&				&	&	&			&			&145.5598	&	&1	&+1\\
\\ \hspace*{\fill} \\			
			&			&120.9461	&				&1	&-1	&			&			&146.1452	&	&1	&-1\\
			&			&			&0.3315			&	&	&			&			&			&0.3371	&&\\
7			&$f_{22}$	&121.2776	&				&1	&0	&20			&$f_{36}$	&146.4823	&	&1	&+0\\
			&			&			&0.3453			&	&	&			&			&			&0.3307	&&\\
			&			&121.6229	&				&1	&+1	&			&			&146.8130	&	&1	&+1\\
\\ \hspace*{\fill} \\			
			&			&121.9554	&				&1	&-1	&			&			&147.6738	&	&1	&-1\\
			&			&			&0.2894			&	&	&			&			&			&0.3667	&&\\
8			&$f_{23}$	&122.2448	&				&1	&0	&21			&$f_{37}$  	&148.0405	&	&1	&0\\
			&			&			&0.2262			&	&	&			&			&			&0.3752	&&\\
			&			&122.4710	&				&1	&+1	&			&			&148.4157	&	&1	&+1\\
\\ \hspace*{\fill} \\			
			&			&122.6706	&				&1	&-1	&			&			&154.6305	&	&1	&-1\\
			&			&			&0.2251			&	&	&			&			&			&0.3542	&&\\
9			&$f_{24}$   &122.8957	&				&1	&0	&22			&$f_{40}$	&154.9847	&	&1	&0\\
			&			&			&0.2934			&	&	&			&			&			&0.355	&&\\
			&			&123.1891	&				&1	&+1	&			&			&155.3397	&	&1	&+1\\	
\\ \hspace*{\fill} \\			
			&			&123.4971	&				&1	&-1	&			&			&156.1622	&	&1	&-1\\
			&			&			&0.3525			&	&	&			&			&			&0.2662	&&\\
10			&$f_{25}$	&123.8496	&				&1	&0	&23			&$f_{41}$	&156.4284	&	&1	&0\\ 
			&			&			&0.2839			&	&	&			&			&			&0.1761	&&\\
			&			&124.1335	&				&1	&+1	&			&			&156.6045	&	&1	&+1\\
\\ \hspace*{\fill} \\			
			&			&124.6147	&				&1	&-1	&			&			&157.1307	&	&1	&-1\\
11			&			&			&0.2903			&	&	&			&			&			&0.2669	&&\\
			&$f_{26}$	&124.9050	&				&1	&0	&24			&$f_{42}$	&157.3976	&	&1	&0\\	
			&			&			&				&	&	&			&			&			&0.2975	&&\\
			&			&			&				&	&	&			&			&157.6951	&	&1	&+1\\	
\\ \hspace*{\fill} \\			
			&			&132.2029	&				&1	&-1	&			&			&158.7194	&	&1	&-1\\
			&			&			&0.57			&	&	&			&			&			&0.359	&&\\
12			&$f_{28}$	&132.5599	&				&1	&0	&25			&$f_{43}$	&159.0784	&	&1	&0\\
			&			&			&0.3462			&	&	&			&			&			&0.3622	&&\\
			&			&132.9061	&				&1	&+1	&			&			&159.4406	&	&1	&+1\\
\\ \hspace*{\fill} \\	
			&			&133.4022	&				&1	&-1\\
			&			&			&0.2757			&	&\\
13			&$f_{29}$	&133.6779	&				&1	&0\\
			&			&			&0.2016			&	&\\
			&			&133.8795	&				&1	&+1\\

\end{longtable}

\begin{table*}
\caption{\label{t7}Model Parameters Comparison for KIC 11968334.}
\centering
\begin{tabular}{lcc}
\hline
\hline
&Input Parameters\\
\hline
&this work &\citet{Triana2017}\\
Mass (M/$M_{\odot}$)                  &1.31          &1.11\\
Initial Metalicity                    &0.024($Z$)    &-\\
                                      &-             &-0.1($[\rm Fe/H]$)\\
Overshooting parameter ($f_{ov}$)     &0.009         &0.0285\\
Mixing length parameter ($\alpha$)    &2.0           &1.9\\
\hline
&Fundamental Parameters\\
\hline
&this work &\citet{Triana2017}\\
Effective temperature ($\rm T_{eff}$/K)  &4782  &4831\\
Radius (R/$R_{\odot}$)                   &5.59  &5.246\\
Large frequency separation ($\Delta\nu$/$\mu$Hz) &11.302    &11.66\\
Period spacing ($\Delta\Pi_{1}$/s)               &78.07     &78.60\\
Core Rotation Rate ($\mu$Hz)         \\
($\delta(\zeta_{mod})$)                     &0.7409$\pm$0.0113   &0.7295$\pm$0.0222  \\
($\delta(\zeta_{as})$)      &0.7300$\pm$0.0110   &0.7430$\pm$0.02566 \\
Envelope Rotation Rate ($\mu$Hz)     \\
($\delta(\zeta_{mod})$)                     &0.0807$\pm$0.0285   &0.0742$\pm$0.0281  \\
($\delta(\zeta_{as})$)      &0.05801$\pm$0.0364  &0.0490$\pm$0.0341   \\
\hline
\end{tabular}
\end{table*}

\section{Conclusion}\label{sec5}

In this study, we construct asteroseismic models for four \textit{Kepler} red giants. Surface effect corrections proposed by \citet{2014A&A...568A.123B} are ineffective for high-frequency $l=0$ and $l=2$ modes. In contrast, g-dominated mixed modes show minimal sensitivity to these effects, confirming their reliability as seismic diagnostics. We therefore use g-dominated mixed modes as constraints to build models that reproduce the observed oscillation frequencies with high accuracy. Applying this approach 
to KIC~9145955, KIC~9970396, KIC~9882316, and KIC~11968334, we determine their fundamental parameters and place constraints on   
their internal structures. The main results are summarized below.  

(1) The fundamental parameters of all four stars are determined by asteroseismic models (see Table \ref{t3}). For KIC 11968334, we also measured the core rotation rate, obtaining a value of $0.7409\pm0.0113 \mu$Hz from its g-dominated mixed modes.

(2) The derived masses and best-fitting convective core overshooting parameters for the four targets, listed in order of increasing stellar mass, are as follows: $M=1.14~M_{\odot}$, $f_{ov}=0.005$ for KIC~9970396; $M=1.20~M_{\odot}$, $f_{ov}=0.007$ for KIC~9145955; $M=1.31~M_{\odot}$, $f_{ov}=0.009$ for KIC~11968334; and $M=1.40~M_{\odot}$, $f_{ov}=0.015$ for KIC~9882316. Although our sample is limited in size, the derived values exhibit a systematic increase of $f_{ov}$ with stellar mass, consistent with the mass dependence reported in previous studies. Our findings suggest that red giants with masses above approximately 1.4 $M_{\odot}$ may require a convective overshooting parameter $f_{ov}$ larger than 0.01 for reliable modeling, which warrants further verification with a larger asteroseismic sample. 

(3) We test the impact of the mixing-length parameter $\alpha$ using KIC 9882316 and find that the fundamental parameters are nearly identical whether all oscillation frequencies or only g-dominated mixed modes are used as constraints. This test obtains a value of $\alpha = 2.05 \pm 0.18$ (using g-dominated mixed modes), which is consistent with our adopted $\alpha = 2.0$. We therefore recommend adopting $\alpha \approx$2.0, which is larger than the solar-calibrated value, for future modeling of red giant stars.

\begin{acknowledgments}
This work is supported by the National Key R\&D Program of China No. 2024YFA1611901, NSFC grants (12090040, 12090042), and the CMS-CSST-2025-A15. This work is also supported by the Joint Research Fund in Astronomy (U2031203) under cooperative agreement between the National Natural Science Foundation of China (NSFC) and the Chinese Academy of Sciences (CAS), the Fundamental Research Funds for the Central Universities. This work is also supported by the Postdoctoral Fellowship Program of CPSF under Grant Number GZC20240126 and the China Postdoctoral Science Foundation under Grant Number 2024M760241. JXC also acknowledges the support of the Postdoctoral Fellowship Program of CPSF under grant Number GZC20240124 and the China Postdoctoral Science Foundation under Grant Number 2024M760242.
X-HC also sincerely appreciates the supports of the Yunnan Revitalization Talent Support Program Young Talent Project, and the Youth Innovation Promotion Association of the Chinese Academy of Science. The authors acknowledge the NASA and Kepler team for allowing them to work with and analyze the Kepler data. The Kepler Mission is funded by NASA’s Science Mission Directorate.
\end{acknowledgments}



\bibliography{zxyref}
\bibliographystyle{aasjournalv7}



\appendix
This is supplementary material to the paper, and it contains four tables and four figures. The observation frequencies and their corresponding observation errors of KIC 9145955, KIC 9970396, KIC 9882316, and KIC 11968334 are listed in Table \ref{ta1}, \ref{ta2}, \ref{ta3}, and \ref{ta4}, respectively. The p-dominated mixed modes identified by us are denoted in these tables. 

The internal structures of the best-fitting models for KIC 9145955, KIC 9970396, KIC 9882316, and KIC 11968334 are illustrated in Figures \ref{fa1}-\ref{fa4}. Panel (a) presents the propagation diagrams of the four red giants, which describes the regions where pressure and gravity waves propagate. Panel (b) shows the determination of the convective envelope, where the convective boundary is identified according to the Schwarzschild criterion, i.e., where the radiative temperature gradient ($\nabla_{r}$) exceeds the adiabatic gradient ($\nabla_{a}$). Panel (c) demonstrates that the stellar density profile exhibits a sharp change in slope at the boundary of the helium core. Panel (d) illustrates that the CNO-cycle reaction rate peaks near this boundary, coinciding with the location of the hydrogen-burning shell.

\renewcommand{\thetable}{\thesection A.\arabic{table}}
\setcounter{table}{0}

\setcounter{figure}{0}
\renewcommand{\thefigure}{A.\arabic{figure}}

\begin{table}[ht!]
\caption{\label{ta1}Oscillation frequencies of KIC 9145955 extracted by \citet{2018ApJ...855...16Z}. The frequency ID is defined by us. The p-dominated mixed modes are marked in the last column.}
\centering
\begin{tabular}{cccccccccc}
\hline\hline
$l$   &ID       &Frequency  &Uncertainty   &Remark   &$l$  &ID      &Frequency  &Uncertainty    &Remark      \\
      &         &($\mu$Hz)  &($\mu$Hz)     &         &     &        &($\mu$Hz)  &($\mu$Hz)      &           \\
\hline
0	&$f_{1}$	&101.811	&0.032	&        	     &1	&$f_{30}$	&125.438	&0.161	&      \\
0	&$f_{2}$	&112.601	&0.012	&	             &1	&$f_{31}$	&126.631	&0.012	&          \\
0	&$f_{3}$	&123.564	&0.012	&	             &1	&$f_{32}$	&127.815	&0.194	&          \\
0	&$f_{4}$	&134.501	&0.090	&                &1	&$f_{33}$	&128.828	&0.0102	&p-dominated         \\
0	&$f_{5}$	&145.568	&0.012	&                &1	&$f_{34}$	&129.573	&0.0102	&         \\
0	&$f_{6}$	&156.749	&0.012	&                &1	&$f_{35}$	&130.624	&0.231	&          \\
0	&$f_{7}$	&168.124	&0.231	&                &1	&$f_{36}$	&131.881	&0.012	&          \\
1	&$f_{8}$	&95.797	    &0.012	&                &1	&$f_{37}$	&135.928	&0.012	&           \\
1	&$f_{9}$	&96.397	    &0.045	&p-dominated     &1	&$f_{38}$	&137.321	&0.088	&           \\
1	&$f_{10}$	&96.931	    &0.029	&                &1	&$f_{39}$	&138.703	&0.231	&              \\
1	&$f_{11}$	&97.463	    &0.231	&                &1	&$f_{40}$	&139.846	&0.231	&p-dominated              \\
1	&$f_{12}$	&98.125	    &0.012	&	             &1	&$f_{41}$	&140.676	&0.154	&        \\
1	&$f_{13}$	&98.840	    &0.012	&                &1	&$f_{42}$	&141.996	&0.012	&          \\
1	&$f_{14}$	&104.314	&0.036	&                &1	&$f_{43}$	&143.495	&0.231	&         \\
1	&$f_{15}$	&105.958	&0.012	&                &1	&$f_{44}$	&148.276	&0.026	&         \\
1	&$f_{16}$	&106.726	&0.113	&              	 &1	&$f_{45}$	&149.881	&0.085	&         \\
1	&$f_{17}$	&107.339	&0.012	&p-dominated     &1	&$f_{46}$	&151.070	&0.051	&p-dominated         \\
1	&$f_{18}$	&107.967	&0.012	&              	 &1	&$f_{47}$	&152.070	&0.044	&         \\
1	&$f_{19}$	&108.778	&0.093	&              	 &1	&$f_{48}$	&153.713	&0.012	&            \\
1	&$f_{20}$	&110.574	&0.012	&              	 &1	&$f_{49}$	&161.014	&0.031	&             \\
1	&$f_{21}$	&113.415	&0.012	&                &1	&$f_{50}$	&162.322	&0.231	&p-dominated             \\
1	&$f_{22}$	&114.413	&0.143	&           	 &1	&$f_{51}$	&163.511	&0.012	&            \\
1	&$f_{23}$	&115.408	&0.168	&                &2	&$f_{52}$	&89.702	    &0.012	&             \\
1	&$f_{24}$	&116.409	&0.012	&                &2	&$f_{53}$	&100.456	&0.012	&             \\
1	&$f_{25}$	&117.368	&0.056	&           	 &2	&$f_{54}$	&111.077	&0.012	&                \\
1	&$f_{26}$	&118.136	&0.012	&p-dominated     &2	&$f_{55}$	&122.039	&0.012	&                 \\
1	&$f_{27}$	&118.842	&0.036	&                &2	&$f_{56}$	&133.199	&0.012	&                 \\
1	&$f_{28}$	&119.787	&0.231	&                &2	&$f_{57}$	&144.186	&0.231	&                 \\
1	&$f_{29}$	&120.864	&0.231	&                &2	&$f_{58}$	&155.474	&0.110	&                 \\
\hline
\end{tabular}
\end{table}

\begin{table}[ht!]
\caption{\label{ta2}Oscillation frequencies of KIC 9970396 extracted by \citet{2018MNRAS.475..981L}. The frequency ID is defined by us. The p-dominated mixed modes are marked in the last column.}
\centering
\begin{tabular}{ccccccccccc}
\hline\hline
$l$   &ID       &Frequency  &Uncertainty   &Remark   &$l$  &ID      &Frequency  &Uncertainty    &Remark      \\
      &         &($\mu$Hz)  &($\mu$Hz)     &         &     &        &($\mu$Hz)  &($\mu$Hz)      &           \\
\hline
0     &$f_{1}$  &45.255     &0.032  &                &1    &$f_{21}$   &66.680   &0.020     &    \\
0     &$f_{2}$  &51.138     &0.029  &                &1    &$f_{22}$   &66.910   &0.040     &   \\
0     &$f_{3}$  &57.396     &0.015  &                &1    &$f_{23}$   &67.120   &0.020     &p-dominated  \\
0     &$f_{4}$  &63.717     &0.023  &                &1    &$f_{24}$   &67.390   &0.023     &  \\
0     &$f_{5}$  &70.031     &0.050  &                &1    &$f_{25}$   &67.680   &0.015     & \\
0     &$f_{6}$  &76.395     &0.077  &                &1    &$f_{26}$   &72.640   &0.070     &\\
1     &$f_{7}$  &48.110     &0.060  &                &1    &$f_{27}$   &73.040   &0.022     &\\
1     &$f_{8}$  &48.350     &0.020  &                &1    &$f_{28}$   &73.290   &0.051     &\\
1     &$f_{9}$  &48.475     &0.034  &                &1    &$f_{29}$   &73.501   &0.019     &p-dominated\\
1     &$f_{10}$ &48.635     &0.016  &                &1    &$f_{30}$   &73.921   &0.041     &\\
1     &$f_{11}$ &54.040     &0.018  &                &1    &$f_{31}$   &79.100   &0.100     &\\
1     &$f_{12}$ &54.195     &0.034  &                &1    &$f_{32}$   &79.477   &0.040     & \\
1     &$f_{13}$ &54.350     &0.016  &p-dominated     &1    &$f_{33}$   &79.850   &0.050     &\\
1     &$f_{14}$ &54.498     &0.020  &                &1    &$f_{34}$   &80.032   &0.032     &p-dominated\\
1     &$f_{15}$ &54.703     &0.029  &                &2    &$f_{35}$   &44.351   &0.021     &\\
1     &$f_{16}$ &60.420     &0.110  &                &2    &$f_{36}$   &50.270   &0.187     &\\
1     &$f_{17}$ &60.650     &0.052  &p-dominated     &2    &$f_{37}$   &56.496   &0.023     &\\
1     &$f_{18}$ &60.899     &0.012  &                &2    &$f_{38}$   &62.895   &0.012     &\\ 
1     &$f_{19}$ &61.140     &0.015  &                &2    &$f_{39}$   &69.173   &0.017     &\\
1     &$f_{20}$ &66.400     &0.065  &                &2    &$f_{40}$   &75.626   &0.091     &\\ 
\hline
\end{tabular}
\end{table}

\begin{table}[ht!]
\caption{\label{ta3}Oscillation frequencies of KIC 9882316 extracted by \citet{corsaro2015}. The frequency ID is defined by us. The p-dominated mixed modes are marked in the last column.}
\centering
\begin{tabular}{cccccccccc}
\hline\hline
$l$   &ID       &Frequency  &Uncertainty   &Remark   &$l$  &ID         &Frequency  &Uncertainty   &Remark      \\
      &         &($\mu$Hz)  &($\mu$Hz)     &         &     &           &($\mu$Hz)  &($\mu$Hz)     &           \\
\hline
0	&$f_{1}$	&140.8260	&0.0112     &            &1    &$f_{23}$   &178.6684   &0.0003        &\\
0   &$f_{2}$    &154.1351	&0.0083     &            &1    &$f_{24}$   &183.7659   &0.0017        &\\
0   &$f_{3}$    &167.7997	&0.0096		&            &1    &$f_{25}$   &186.3344   &0.0004        &\\
0   &$f_{4}$    &181.3776	&0.0049		&            &1    &$f_{26}$   &188.1812   &0.0042        &p-dominated \\
0   &$f_{5}$    &194.9649	&0.0075		&            &1    &$f_{27}$   &189.8208   &0.0026        &\\
0   &$f_{6}$    &208.7613	&0.0225		&            &1    &$f_{28}$   &192.4676   &0.0004        &\\
0   &$f_{7}$    &222.7163	&0.0217		&            &1    &$f_{29}$   &195.3718   &0.0007        &\\
1   &$f_{8}$    &145.1127	&0.0011     &            &1    &$f_{30}$   &198.3582   &0.0024        &\\
1   &$f_{9}$    &146.6253	&0.0002     &            &1    &$f_{31}$   &201.0854   &0.0053        &\\ 
1   &$f_{10}$   &147.6557	&0.0054		&p-dominated &1    &$f_{32}$   &202.6380   &0.0087        &p-dominated   \\
1   &$f_{11}$   &148.9356	&0.0083     &            &1    &$f_{33}$   &205.2324   &0.0012        &\\
1   &$f_{12}$   &150.6122	&0.0012     &            &1    &$f_{34}$   &211.8795   &0.0012        &\\
1   &$f_{13}$   &156.1389	&0.0016     &            &1    &$f_{35}$   &214.9523   &0.0137        &\\
1   &$f_{14}$   &158.0426	&0.0016     &            &1    &$f_{36}$   &216.7064   &0.0143        &p-dominated \\
1   &$f_{15}$   &159.8878	&0.0007     &            &1    &$f_{37}$   &219.6626   &0.0013        &\\
1   &$f_{16}$   &161.1611	&0.0088		&p-dominated &2    &$f_{38}$   &139.1263   &0.0190        &\\
1   &$f_{17}$   &162.5396	&0.0036     &            &2    &$f_{39}$   &152.4045   &0.0218        &\\
1   &$f_{18}$   &164.5319	&0.0004     &            &2    &$f_{40}$   &166.0895   &0.0086        &\\
1   &$f_{19}$   &171.1069	&0.0009     &            &2    &$f_{41}$   &179.7663   &0.0068        &\\
1   &$f_{20}$   &173.2846	&0.0002     &            &2    &$f_{42}$   &193.4584   &0.0142        &\\
1   &$f_{21}$   &174.7567	&0.0044     &p-dominated &2    &$f_{43}$   &207.2872   &0.0162        &\\
1   &$f_{22}$   &176.3385	&0.0002     &            &2    &$f_{44}$   &221.2658   &0.0300        &\\
\hline
\end{tabular}
\end{table}

\begin{table}[ht!]
\caption{\label{ta4}Oscillation frequencies of KIC 11968334 extracted by \citet{corsaro2015}. The frequency ID is defined by us. The p-dominated mixed modes are marked in the last column.}
\centering
\begin{tabular}{cccccccccccc}
\hline\hline
$l$ &m  &ID         &Frequency  &Uncertainty   &Remark  &$l$ &m &ID   &Frequency  &Uncertainty   &Remark      \\
    &   &           &($\mu$Hz)  &($\mu$Hz)     &        &    &        &($\mu$Hz)  &($\mu$Hz)     &           \\
\hline
0	&0	&$f_{1}$	&105.8760	&0.0124	   &            &1	 &-1     &$f_{44}$  &134.1877   &0.0042 &        \\
0	&0	&$f_{2}$	&116.7790	&0.0068	   &            &1	 &0	     &$f_{45}$	&134.4257	&0.0026	&         \\  
0	&0	&$f_{3}$	&128.0901	&0.0051	   &            &1   &+1     &$f_{46}$   &134.6927  &0.0002 &          \\
0	&0	&$f_{4}$	&139.4736	&0.0023	   &            &1   &-1     &$f_{47}$	&135.2455   &0.0036 &          \\
0	&0	&$f_{5}$	&150.7704	&0.0045	   &            &1	 &0	     &$f_{48}$	&135.5818	&0.0003	&           \\
0	&0	&$f_{6}$	&162.3268	&0.0054	   &            &1	 &0	     &$f_{49}$	&136.9763	&0.0006	&           \\
0	&0	&$f_{7}$	&174.0532	&0.0156	   &            &1	 &+1	 &$f_{50}$	&137.3733	&0.0008	&           \\   
1	&0	&$f_{8}$	&100.2691	&0.0014	   &            &1	 &-1	 &$f_{51}$	&142.5812	&0.0003 &           \\
1	&0	&$f_{9}$	&100.5094	&0.0009	   &            &1	 &0	     &$f_{52}$	&142.9513	&0.0003	&           \\
1	&0	&$f_{10}$	&101.3712	&0.0012	   &            &1   &+1     &$f_{53}$	&143.2594   &0.0008	&           \\
1	&0	&$f_{11}$	&110.7372	&0.0010	   &            &1	 &-1	 &$f_{54}$	&144.0835  	&0.0002	&           \\
1	&+1	&$f_{12}$	&111.0597	&0.0018	   &            &1	 &0	     &$f_{55}$	&144.4057	&0.0001	&           \\  
1	&-1	&$f_{13}$	&111.2019	&0.0007	   &            &1	 &+1	 &$f_{56}$	&144.7047	&0.0001	&            \\
1	&0	&$f_{14}$	&111.4316	&0.0043	   &p-dominated &1	 &-1	 &$f_{57}$	&145.1983	&0.0038	&             \\
1	&+1	&$f_{15}$	&111.7049	&0.0079	   &            &1	 &0	     &$f_{58}$	&145.3729	&0.0022	&p-dominated   \\   
1	&0	&$f_{16}$	&112.0552	&0.0041	   &            &1	 &+1	 &$f_{59}$	&145.5598	&0.0025	&             \\
1	&+1	&$f_{17}$	&112.3600	&0.0094	   &            &1	 &-1	 &$f_{60}$  &146.1452	&0.0002	&           \\ 
1	&0	&$f_{18}$	&112.9492	&0.0013	   &            &1	 &0	     &$f_{61}$	&146.4823	&0.0016	&           \\
1	&+1	&$f_{19}$	&113.2798	&0.0012	   &            &1	 &-1	 &$f_{62}$ 	&147.6738	&0.0006	&            \\
1	&-1	&$f_{20}$	&118.6890	&0.0016	   &            &1	 &0	     &$f_{63}$	&148.0405	&0.0007	&            \\
1	&0	&$f_{21}$	&118.9324	&0.0017	   &            &1	 &+1	 &$f_{64}$	&148.4157	&0.0005	&             \\
1	&-1	&$f_{22}$	&119.8195	&0.0014	   &            &1	 &0	     &$f_{65}$	&151.4637	&0.0011	&             \\
1	&0	&$f_{23}$	&120.0887	&0.0012	   &            &1	 &0	     &$f_{66}$	&153.2149	&0.0009	&            \\
1	&-1	&$f_{24}$	&120.9461	&0.0105	   &            &1	 &-1	 &$f_{67}$	&154.6305	&0.0003	&           \\
1	&0	&$f_{25}$	&121.2776	&0.0014	   &            &1	 &0	     &$f_{68}$	&154.9847	&0.0003	&              \\
1	&+1	&$f_{26}$	&121.6229	&0.0074	   &            &1	 &+1	 &$f_{69}$	&155.3397	&0.0007	&             \\
1	&-1	&$f_{27}$	&121.9554	&0.0062	   &            &1	 &-1	 &$f_{70}$	&156.1622	&0.0077	&             \\
1	&0	&$f_{28}$	&122.2448	&0.0040	   &p-dominated &1	 &0	     &$f_{71}$	&156.4284	&0.0049	&p-dominated   \\ 
1	&+1	&$f_{29}$	&122.4710	&0.0007	   &            &1	 &+1	 &$f_{72}$	&156.6045	&0.0081	&             \\
1	&-1	&$f_{30}$	&122.6706	&0.0076	   &            &1	 &-1	 &$f_{73}$	&157.1307	&0.0068	&              \\
1	&0	&$f_{31}$	&122.8957	&0.0004	   &            &1	 &0	     &$f_{74}$	&157.3976	&0.0039	&            \\
1	&+1	&$f_{32}$	&123.1891	&0.0018	   &            &1	 &+1	 &$f_{75}$	&157.6951	&0.0113	&            \\ 
1	&-1	&$f_{33}$	&123.4971	&0.0013	   &            &1	 &-1	 &$f_{76}$	&158.7194	&0.0005	&            \\
1	&0	&$f_{34}$	&123.8496	&0.0009	   &            &1	 &0	     &$f_{77}$	&159.0784	&0.0010	&            \\
1	&+1	&$f_{35}$	&124.1335	&0.0010	   &            &1	 &+1	 &$f_{78}$	&159.4406	&0.0007	&            \\
1	&-1	&$f_{36}$	&124.6147	&0.0007	   &            &2	 &0	     &$f_{79}$	&104.4480	&0.0121	&             \\
1	&0	&$f_{37}$	&124.9050	&0.0009	   &            &2	 &0	     &$f_{80}$	&115.2533	&0.0071	&              \\
1	&-1	&$f_{38}$	&132.2029	&0.0005	   &            &2	 &0	     &$f_{81}$	&126.6288	&0.0055	&             \\
1	&0	&$f_{39}$	&132.5599	&0.0002	   &            &2	 &0	     &$f_{82}$	&138.0535	&0.0047	&              \\
1	&+1	&$f_{40}$	&132.9061	&0.0003	   &            &2	 &0	     &$f_{83}$	&149.3991	&0.0039	&             \\
1	&-1	&$f_{41}$	&133.4022	&0.0002	   &            &2	 &0	     &$f_{84}$	&161.0338	&0.0164	&             \\
1	&0	&$f_{42}$	&133.6779	&0.0003	   &p-dominated &2	 &0	     &$f_{85}$	&172.7089	&0.0164	&             \\
1	&+1	&$f_{43}$	&133.8795	&0.0024	   &            &	 &	     &		    &		    &	    &              \\
\hline
\end{tabular}
\end{table}

\begin{figure}[ht!]
\centering
\includegraphics[angle=0,scale=1.2]{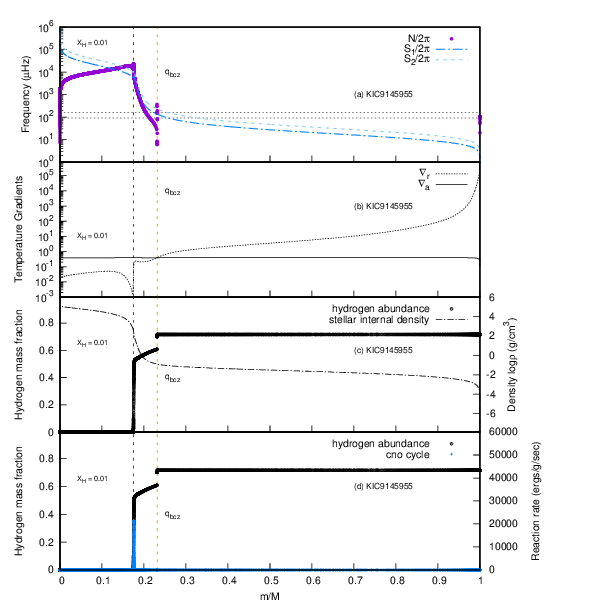}
\caption{Internal structure of KIC 9145955. The vertical red dashed lines in panel (a) to panel (d) denote the helium core boundary where the hydrogen mass fraction is about 0.01. The vertical orange dashed lines in panel (a) to panel (d) denote the base of the convection zone, which is marked as $q_{bcz}$. In panel (a), two horizontal dashed lines represent the range of observed frequencies. The purple dot indicates the Brunt-V$\ddot{\rm a}$is$\ddot{\rm a}$l$\ddot{\rm a}$ frequency $N$ and the blue dash-dotted and light-blue dashed line indicates characteristic acoustic frequency $S_{1}$ and $S_{2}$, respectively. In panel (b), the continues line and dotted line indicate adiabatic and radiative temperature gradients. In panel (c), circles and a dotted line indicate the distribution of the hydrogen abundance and inner density. In panel (d), blue crosses represent the reaction rate of the carbon, nitrogen and oxygen (CNO) cycle.}
\label{fa1}
\end{figure}

\begin{figure}[ht!]
\centering
\includegraphics[angle=0,scale=1.2]{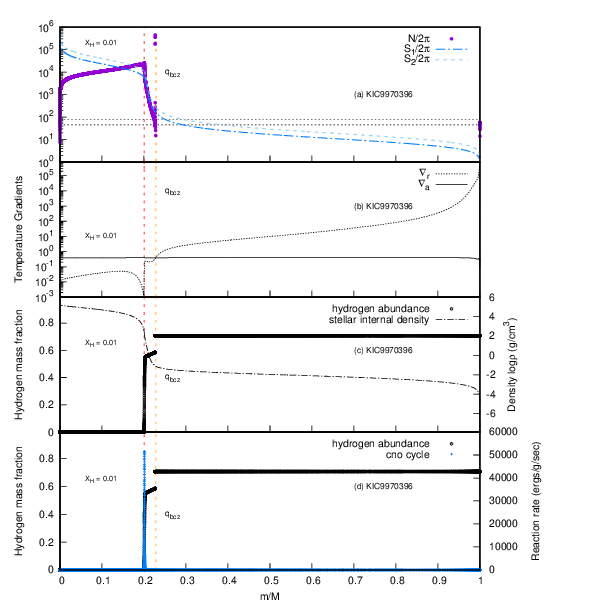}
\caption{Internal structure of KIC 9970396. The vertical red dashed lines in panel (a) to panel (d) denote the helium core boundary where the hydrogen mass fraction is about 0.01. The vertical orange dashed lines in panel (a) to panel (d) denote the base of the convection zone, which is marked as $q_{bcz}$. In panel (a), two horizontal dashed lines represent the range of observed frequencies. The purple dot indicates the Brunt-V$\ddot{\rm a}$is$\ddot{\rm a}$l$\ddot{\rm a}$ frequency $N$ and the blue dash-dotted and light-blue dashed line indicates characteristic acoustic frequency $S_{1}$ and $S_{2}$, respectively. In panel (b), the continues line and dotted line indicate adiabatic and radiative temperature gradients. In panel (c), circles and a dotted line indicate the distribution of the hydrogen abundance and inner density. In panel (d), blue crosses represent the reaction rate of the carbon, nitrogen and oxygen (CNO) cycle.}
\label{fa2}
\end{figure}

\begin{figure}[ht!]
\centering
\includegraphics[angle=0,scale=1.2]{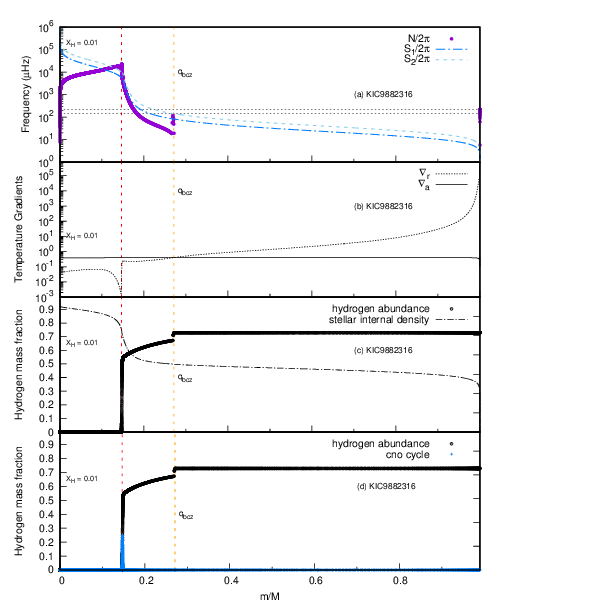}
\caption{Internal structure of KIC 9882316. The vertical red dashed lines in panel (a) to panel (d) denote the helium core boundary where the hydrogen mass fraction is about 0.01. The vertical orange dashed lines in panel (a) to panel (d) denote the base of the convection zone, which is marked as $q_{bcz}$. In panel (a), two horizontal dashed lines represent the range of observed frequencies. The purple dot indicates the Brunt-V$\ddot{\rm a}$is$\ddot{\rm a}$l$\ddot{\rm a}$ frequency $N$ and the blue dash-dotted and light-blue dashed line indicates characteristic acoustic frequency $S_{1}$ and $S_{2}$, respectively. In panel (b), the continues line and dotted line indicate adiabatic and radiative temperature gradients. In panel (c), circles and a dotted line indicate the distribution of the hydrogen abundance and inner density. In panel (d), blue crosses represent the reaction rate of the carbon, nitrogen and oxygen (CNO) cycle.}
\label{fa3}
\end{figure}

\begin{figure}[ht!]
\centering
\includegraphics[angle=0,scale=1.2]{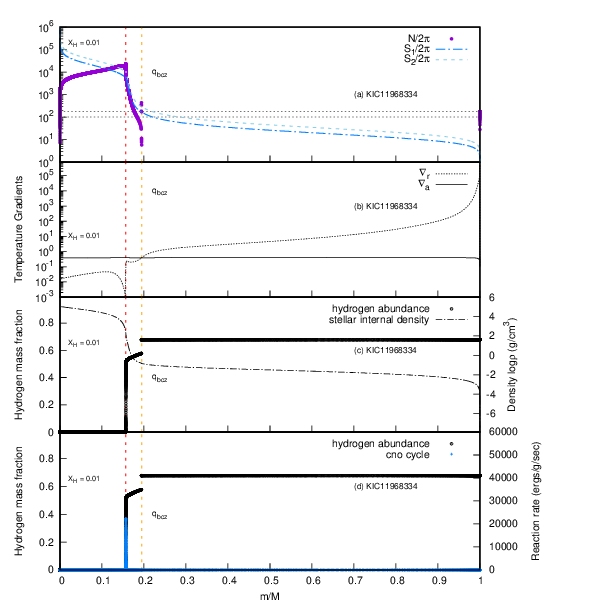}
\caption{Internal structure of KIC 11968334. The vertical red dashed lines in panel (a) to panel (d) denote the helium core boundary where the hydrogen mass fraction is about 0.01. The vertical orange dashed lines in panel (a) to panel (d) denote the base of the convection zone, which is marked as $q_{bcz}$. In panel (a), two horizontal dashed lines represent the range of observed frequencies. The purple dot indicates the Brunt-V$\ddot{\rm a}$is$\ddot{\rm a}$l$\ddot{\rm a}$ frequency $N$ and the blue dash-dotted and light-blue dashed line indicates characteristic acoustic frequency $S_{1}$ and $S_{2}$, respectively. In panel (b), the continues line and dotted line indicate adiabatic and radiative temperature gradients. In panel (c), circles and a dotted line indicate the distribution of the hydrogen abundance and inner density. In panel (d), blue crosses represent the reaction rate of the carbon, nitrogen and oxygen (CNO) cycle.}
\label{fa4}
\end{figure}

\end{CJK*}
\end{document}